\documentclass[conference,compsoc]{IEEEtran}
\ifCLASSOPTIONcompsoc
  \usepackage[nocompress]{cite}
\else
  \usepackage{cite}
\fi
\ifCLASSINFOpdf
  \usepackage[pdftex]{graphicx}
\else
  \usepackage[dvips]{graphicx}
\fi
\usepackage{amsmath}
\usepackage{algorithm}
\usepackage{algpseudocode}
\usepackage{setspace} 

\usepackage{array}

\ifCLASSOPTIONcompsoc
 \usepackage[caption=false,font=footnotesize,labelfont=sf,textfont=sf]{subfig}
\else
 \usepackage[caption=false,font=footnotesize]{subfig}
\fi

\usepackage{wasysym}
\usepackage{pifont}

\usepackage{ragged2e}
\usepackage{amssymb}

\usepackage{booktabs}
\usepackage{multirow}
\usepackage[table]{xcolor}
\usepackage{makecell}

\usepackage{tcolorbox}

\usepackage [english]{babel}
\usepackage [autostyle, english = american]{csquotes}
\MakeOuterQuote{"}

\newcolumntype{C}[1]{>{\centering\arraybackslash}p{#1}}
\newcolumntype{L}[1]{>{\raggedright\arraybackslash}p{#1}}
\definecolor{lightgray}{gray}{0.9}

\begin{document}

%
\title{SoK: Decoding the Enigma of Encrypted Network Traffic Classifiers}






%
\author{

\IEEEauthorblockN{
Nimesha Wickramasinghe\IEEEauthorrefmark{1},
Arash Shaghaghi\IEEEauthorrefmark{1},
Gene Tsudik\IEEEauthorrefmark{2}, 
Sanjay Jha\IEEEauthorrefmark{1}
}

\IEEEauthorblockA{
\IEEEauthorrefmark{1}School of Computer Science and Engineering, The University of New South Wales, Sydney, Australia. }


\IEEEauthorblockA{\IEEEauthorrefmark{2}School of Information \& Computer Sciences, University of California Irvine, USA}



}


\maketitle

%

\begin{abstract}
The adoption of modern encryption protocols such as TLS 1.3 has significantly challenged traditional network traffic classification (NTC) methods. As a consequence, researchers are increasingly turning to machine learning (ML) approaches to overcome these obstacles. This paper analyses ML-based NTC studies by developing a taxonomy of their design choices, benchmarking suites, and prevalent assumptions impacting classifier performance. Through this systematization, we demonstrate widespread reliance on outdated datasets, oversights in design choices, and the consequences of unsubstantiated assumptions. Our evaluation reveals that the majority of proposed encrypted traffic classifiers have mistakenly utilized unencrypted traffic due to the use of legacy datasets. Furthermore, by conducting 348 feature occlusion experiments on state-of-the-art classifiers, we show how oversights in NTC design choices lead to overfitting and validate or refute prevailing assumptions with empirical evidence. By highlighting lessons learned, we offer strategic insights, identify emerging research directions, and recommend best practices to support the development of real-world applicable NTC methodologies.
\end{abstract}

\section{Introduction}
Network Traffic Classification (NTC) is a fundamental process that identifies and categorizes traffic into predefined classes. NTC is crucial for various applications, including network management and security, Quality of Service (QoS) provisioning, and lawful interception. However, the increasing adoption of encryption protocols, particularly Transport Layer Security (TLS) 1.3, has introduced significant challenges to traditional NTC methods. Encrypted traffic obscures payload content, rendering many conventional classification techniques ineffective and necessitating new approaches that do not rely on signatures in plain-text payloads (e.g., deep packet inspection \cite{dpi_survey}). As a result, researchers are increasingly turning to machine learning (ML) to address these challenges \cite{foundation_et_bert, foundation_yatc_full}.

Despite the pressing need for effective NTC in the context of modern encrypted traffic, current research efforts face several challenges. One of them is the reliance on outdated datasets collected before 2018 \cite{datasets_iscx_tor, datasets_iscx_vpn, datasets_ustc_tfc, datasets_cross_platform}. We show that these legacy datasets do not accurately reflect the characteristics of contemporary network traffic, particularly with the adoption of TLS 1.3. Furthermore, such datasets often contain unencrypted traffic or utilize deprecated cipher suites (e.g., 3DES and RC4), leading to deceptive results when training and evaluating ML models.

Another significant challenge stems from design choices when developing NTC models. Many studies overlook the potential for overfitting due to session-specific artifacts, which are often uninformative (i.e., initialized using pseudo-random values during session establishment). We show that this results in classifiers that perform well on test data but fail to generalize to real-world scenarios, undermining their robustness and reliability.

Furthermore, the field is fraught with unsubstantiated and often conflicting assumptions. Many studies assume that imperfect randomness of ciphers causes discernible patterns in encrypted payloads, which can be exploited for NTC \cite{encrypted_network_traffic_classification_using, deep_packet, cbd, foundation_et_bert, bfcn, foundation_flow_mae, foundation_yatc_full}. By making this assumption increasingly questionable, TLS 1.3 guarantees that the only learnable characteristic of a ciphertext is its length \cite{tls_13}. Additionally, there are disputes over the impact of practices such as truncating or padding payload data on classification performance \cite{markov-gan}. These conflicting views create confusion and hinder the development of effective NTC methodologies.

To the best of our knowledge, these methodological pitfalls across the entire classification pipeline have not received sufficient scrutiny. Therefore, we address outstanding challenges in NTC through this study by making the following contributions:
\begin{enumerate}
    \item \textbf{Systematization of Knowledge}: We comprehensively analyse NTC studies, identifying their design choices and benchmarking suites.

    \item \textbf{Discuss pitfalls in NTC}: By critically evaluating widely used network traffic datasets, design choices, and common assumptions, we identify and demonstrate common pitfalls in NTC. 

    \item \textbf{CipherSpectrum}: We introduce CipherSpectrum, a contemporary network traffic dataset, to address the limitations of existing datasets. For cipher-agnostic NTC, the dataset uniformly presents traffic sessions encrypted with all three mandated/recommended cipher suites of TLS 1.3. 

    \item \textbf{Strategic insights}: Building on lessons learned, we propose best practices and future research directions that lead to accurate, generalizable and real-world applicable network traffic classifiers.
\end{enumerate}

\textbf{Scope of the Paper}: This paper concentrates on raw-information based NTC \cite{foundation_et_bert, foundation_yatc, foundation_flow_mae, deep_packet}, which extracts information directly from network traffic, such as packet headers and payloads. While we exclude side-channel based approaches \cite{stat_1, stat_2, stat_3, stat_4} and multimodal methods \cite{multi_1, multi_2, multi_3, multi_4}, we recognize that a thorough understanding of raw data is fundamental to NTC. By focusing on how raw data can be meticulously used to classify traffic, we aim to provide insights that can enhance raw information-based classifiers and inform and improve multimodal approaches. The increasing adoption of raw-information-based NTC \cite{raw_info_popularity} further underscores the relevance and timeliness of this study.

This paper is organized as follows. In Section \ref{sec:taxonomy_and_classification}, we present a taxonomy and classification of existing NTC studies, detailing their design choices, benchmarking suites, and prevalent assumptions. Section \ref{sec:systematization} provides a systematization of these studies, synthesizing insights and identifying common trends. In Section \ref{sec:unresolved_challenges}, we outline the unresolved challenges in NTC by formulating specific research questions that address the issues of outdated datasets, design oversights, and unsubstantiated assumptions. Section \ref{sec:decoding_the_ntc_fidelity} presents our empirical investigations, where we prove our conjectures and answer the research questions through extensive experiments. Finally, in Section 6, we discuss the implications of our findings and highlight emerging research directions.

\section{Taxonomy and Classification}
\label{sec:taxonomy_and_classification}
In this section, we lay the groundwork for our network traffic classification (NTC) study using raw information. We introduce a taxonomy that organizes the various aspects of classifier development, focusing on the key choices made throughout the process. Broadly, these decisions fall into two main categories: design choices and benchmarking choices, which are discussed in Section \ref{subsec:preliminries_design_choices} and Section \ref{subsec:preliminries_tasks_and_datasets} respectively. This framework serves as a basis for understanding and analyzing the development and evaluation of network traffic classifiers in this paper.

\subsection{Design Choices}
\label{subsec:preliminries_design_choices}
The design of a Network Traffic Classification (NTC) system involves several ancillary decisions that impact its functionality, accuracy, and suitability for specific tasks. Key design considerations include the granularity of data, the methods used to extract relevant information, and the selection of features for analysis. For instance, MAC addresses, IP addresses, and protocol ports are commonly referred to as Strong Identification Information (SII) \cite{foundation_et_bert}. Incorporating SII constitutes an important design choice within the system’s overall framework.

\begin{figure}[tb]
    \centering
    
    \includegraphics[width=1\columnwidth]{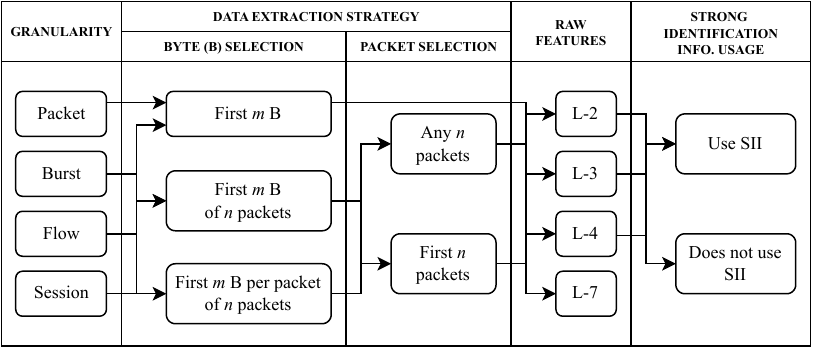}
    
    \caption{Taxonomy of design choices in raw information based NTC}
    \label{img:1_sok_taxonomy}
    
\end{figure}

As illustrated in Figure \ref{img:1_sok_taxonomy}, each choice impacts how the classifier processes network data, which is crucial in refining the classifier's capability. This section delves into these design considerations, structured to reflect the decision-making paths highlighted in Figure \ref{img:1_sok_taxonomy}.

\subsubsection{Traffic Granularity}
\label{subsubsec:traffic_granularity}
Traffic granularity defines the level of detail at which network traffic is analyzed and classified. To formalize, we represent the set of all traffic packets as $P=\left\{p^1,...,p^{|p|}\right\}$, where each packet $p^i=(x^i,y^i,t^i)$ for $i \in [1,|P|]$. In this notation, $x^i$ represents the 5-tuple consisting of the source IP, destination IP, source port, destination port, and protocol; $y^i$ denotes the packet size in bytes, with $y^i \in (0,\infty)$; and $t^i$ stands for the packet's transmission timestamp, measured in seconds, with $t^i \in [0,\infty)$. The most prominent traffic granularities found in the literature include:

\textbf{Packet Granularity} represents the finest level of analysis, where each packet $p^i=(x^i,y^i,t^i)$ is individually examined. 

\textbf{Burst Granularity} analyzes traffic in bursts, or clusters of packets transmitted within short intervals, separated by idle periods. A burst $B^n$ consists of packets that arrive within a specified time window defined by the burst duration threshold $\Delta t$. Formally, let $B$ represent the set of all bursts, and each burst $b^n$ can be defined as: $b^n = \left\{ p^i \in P \mid t^i - t^{i-1} \leq \Delta t \right\}$, where $n\in[1,|B|]$.

\textbf{Flow Granularity} groups all unidirectional packets with the same 5-tuple $x^i$. Let $F$ represent the set of all flows, where each flow $f^k$ is a collection of packets with the same $x^i$. Formally, the flow $f^k$ can be defined as: \\ 
$f^k = \left\{ p^i \in P \mid x^i = x^k\right\}$, where $x^k$ is the 5-tuple identifying the flow $f^k$.

\textbf{Session Granularity} (or \textit{Bi-Flow Granularity}) aggregates traffic flows between two endpoints into a single granularity by treating flows in both directions as part of one conversation. In this approach, the 5-tuple $x^i$ is treated as bi-directional, meaning that flow $A$ $\rightarrow$ $B$ and $B$ $\rightarrow$ $A$ are combined. Formally, let $S$ represent the set of all sessions, where each session $s^m$ consists of paired flows $f^k$ that capture the full interaction between two endpoints

\subsubsection{Data Extraction Strategy}
\label{subsubsec:data_extraction_strategy}
Once the traffic granularity is determined, the next essential design choice involves selecting a data extraction strategy. This strategy defines how data is captured from the chosen granularity level and how it is represented. Referring to Figure \ref{img:1_sok_taxonomy}, we categorize data extraction strategies into the following types:

\textbf{Type 1: First \textit{m} Bytes of Selected Granularity}
This strategy involves extracting the first $m$ bytes from the entire data segment of the selected granularity level, whether it is packet, burst, flow, or session. This approach treats the selected granularity as a continuous data stream without regard to individual packet boundaries or the number of packets it contains. As shown in Figure \ref{img:1_sok_taxonomy}, this method is the default—and, effectively, the only—option available for packet granularity, where each packet's raw information is extracted for classification.

\textbf{Type 2: First \textit{m} Bytes of \textit{n} Packets}
In this method, the classifier extracts the first $m$ bytes collectively from a sequence of $n$ packets within the selected granularity (e.g., burst, flow, or session). If the total size of these $n$ packets is less than $m$, padding is applied to reach the desired byte length. Conversely, if the total size exceeds $m$, it is truncated to match the specified size.

\textbf{Type 3: First \textit{m} Bytes Per Packet of \textit{n} Packets}
This approach extracts the first $m$ bytes from each of the first $n$ packets within the selected granularity, resulting in a total of $m \times n$ bytes. By preserving packet boundaries and treating the initial bytes of each packet separately, this strategy retains the structural integrity of network traffic.

As Figure \ref{img:1_sok_taxonomy} depicts, an additional choice for the \textit{Type 2} and \textit{Type 3} strategies is the selection of n packets from which data will be extracted, offering two primary options:

\textbf{First \textit{n} Packets}: This option selects the initial $n$ packets of the chosen granularity (burst, flow, or session) under the assumption that the earliest packets often contain critical information, such as protocol handshakes or initial data exchanges.

\textbf{Any Consecutive \textit{n} Packets}: Alternatively, the $n$ packets can be any consecutive sequence within the granularity, providing flexibility to capture patterns that may occur at various stages of the communication.

\subsubsection{Raw Features}
Building on the selected traffic granularity and data extraction strategy, the next aspect in designing a network traffic classification (NTC) system is the choice of raw features, as illustrated in Figure \ref{img:1_sok_taxonomy}. Raw features refer to the unprocessed data extracted directly from network traffic without prior aggregation or manipulation. These features play a critical role in NTC as they often capture unique patterns essential for accurate classification. The choice of traffic granularity and data extraction strategy shapes the availability, scope, and type of raw features that can be used. Broadly, these features can be categorized into packet header information and application layer payloads.

Packet headers provide essential metadata about the packet’s path through the network. 

\textbf{Layer-2 (L2)}: The Ethernet layer comprises attributes such as source and destination MAC addresses, which identify devices on the local network. 

\textbf{Layer-3 (L3)}: The Network layer mainly includes attributes like source and destination IP addresses, which are critical for routing data among networks. 

\textbf{Layer-4 (L4)}: The Transport layer primarily consists of source and destination ports, along with other flow control and error recovery features.

\textbf{Layer-7 (L7)}: The Application layer payloads contain the actual data transmitted by applications. However, with the growing prevalence of encryption protocols, much of this payload data is encrypted, limiting accessibility to plaintext for analysis.

\subsubsection{Strong Identification Information (SII)}
\label{subsubsec:sii}
Following the selection of raw features, another critical consideration in designing network traffic classification (NTC) systems is the use of Strong Identification Information (SII), as outlined in Figure \ref{img:1_sok_taxonomy}. SII refers to features directly identifying a user, device, or application, providing highly discriminative information that can significantly enhance classification accuracy. 

According to literature, typical examples of SII include: MAC addresses in L2 (ethernet layer), which are unique identifiers for network interfaces; IP addresses in L3 (network layer), serving as numerical labels assigned to devices on a network; and protocol ports in L4 (transport layer), which are numeric identifiers for specific applications or services.

While some studies opt to obfuscate SII to mitigate overfitting, others choose to incorporate SII to leverage its discriminative power.

\subsection{Downstream Tasks and Benchmarking}
\label{subsec:preliminries_tasks_and_datasets}
Network traffic classification covers a range of downstream tasks focused on identifying specific network activities, applications, or security threats. Benchmarking in this context involves selecting suitable datasets to rigorously test and validate the classifier's effectiveness in performing the chosen downstream task. Several public datasets are available as benchmarks, each addressing a specific classification task.

A significant area of focus as a downstream task is detecting malicious activities, including malware communications, network intrusions, and botnet traffic. The USTC-TFC2016 dataset \cite{datasets_ustc_tfc} provides samples of malicious and benign traffic, enabling the development of classifiers capable of detecting malware within network flows. Similarly, the CIC-IDS2018 dataset \cite{datasets_cic_ids2018} contains a range of intrusion scenarios alongside benign traffic. Additionally, CIC-IoT2022 \cite{datasets_cic_iot2022} and Bot-IoT \cite{datasets_bot_iot} datasets focus on IoT environments, offering traffic data that includes both normal device activities and malicious behaviours associated with botnets.

Another critical task involves classifying encrypted and anonymized traffic, such as VPN and Tor communications. The ISCXVPN2016 dataset \cite{datasets_iscx_vpn} includes traffic data from both VPN and non-VPN connections, assisting researchers in creating models that can identify the use of VPNs. Similarly, the ISCXTor2016 dataset \cite{datasets_iscx_tor} focuses on Tor traffic classification, offering both Tor and non-Tor traffic samples. 

In application identification, web and mobile traffic classification has gained prominence. The CSTNET-TLS1.3 dataset \cite{foundation_et_bert} supports web traffic classification by providing samples of web communications using TLS 1.3. The Cross-Platform Application dataset \cite{datasets_cross_platform} offers network traffic generated by various mobile applications across different platforms, including Android and iOS.

In addition to these public datasets, private datasets collected from operational networks or controlled experiments are often employed to evaluate network traffic classifiers. \\

At its core, ML in NTC transforms raw network data into meaningful insights by identifying patterns and relationships that may not be explicitly defined. The interplay between data granularity, feature selection, and benchmarking choices shapes how models learn and perform, ultimately determining their effectiveness in real-world scenarios.

\section{Systematization}
\label{sec:systematization}

To provide a comprehensive overview of the existing research on Network Traffic Classification (NTC), we conduct a systematic literature survey focusing on studies that utilize raw network traffic information. Our objective is to identify and analyze key literature that discusses the design choices and methodologies pertinent to our scope, particularly in the context of encrypted communications.

\subsection{Methodology}
We initiate the survey by conducting an extensive search across multiple academic databases, including JSTOR, SCOPUS, EBSCO, and Google Scholar. The search is guided by the following query: \textit{("network traffic" OR "encrypted traffic") AND ("classification" OR "analysis" OR "detection") AND ("review" OR "survey" OR "sok")}. This search string was designed to capture a broad spectrum of literature related to NTC, emphasizing works that offer a critical overview or synthesis of the field. Since the first raw-information based network traffic classifier was published in 2015 \cite{mobile_encrypted_traffic_classification, encrypted_network_traffic_classification_using, cbd}, we filtered our search to studies from 2015 onward.

The initial search yields approximately 152 articles. To refine this pool of literature, we exclude studies that were (a) not directly related to the subject matter, such as those focusing on unrelated networking topics or other domains (e.g., road traffic analysis, NTC studies that do not leverage ML); (b) non-peer-reviewed, to ensure the academic rigour and credibility of the sources; and (c) duplicate studies, to maintain a unique set of references. From the refined list of papers, we selected raw-information based NTC studies that detail their design choices and the benchmarking suites used (see Section \ref{sec:taxonomy_and_classification}).

\subsection{Culmination}
The collected articles are then meticulously analyzed and grouped based on the themes and design choices discussed in Section \ref{sec:taxonomy_and_classification}. These categories include traffic granularity, data extraction strategy, raw features used, Strong Identification Information (SII) and dataset usage. We present the culmination of this process in Table \ref{tab:1_lit_summary}, which provides a concise reference to support our subsequent analysis and discussions.


\begin{table*}[!t]
\renewcommand{\arraystretch}{1.2}

    \caption{Raw information based network traffic classification}
    \label{tab:1_lit_summary}
    
    \centering
    \normalsize

    \resizebox{\textwidth}{!}{
        \rowcolors{2}{}{lightgray}        
        \begin{tabular}{cc*{4}{C{.7cm}}*{1}{C{1.5cm}}*{1}{C{1.7cm}}*{4}{C{1.15cm}}*{4}{C{1.1cm}}c*{1}{C{1.1cm}}}

            \toprule

                \multicolumn{2}{c}{\textbf{Study}} &
                \multicolumn{4}{c}{\textbf{Raw Features Used}} &
                \multicolumn{2}{c}{\textbf{Data Extraction}} &
                \multicolumn{4}{c}{\textbf{Traffic Granularity}} &
                \multicolumn{6}{c}{\textbf{Datasets Used}}
                \\

                \cmidrule(lr){1-2} 
                \cmidrule(lr){3-6} 
                \cmidrule(lr){7-8} 
                \cmidrule(lr){9-12}
                \cmidrule(lr){13-18}

                \rowcolor{white}
                \textbf{Ref.} &
                \textbf{Year} &
                
                \textbf{L2} &
                \textbf{L3} &
                \textbf{L4} &
                \textbf{L7} &
                
                \textbf{\# Bytes} &
                \textbf{\# Packets} &

                \textbf{Packet} &
                \textbf{Burst} &
                \textbf{Flow} &
                \textbf{Session} &
                
                \textbf{VPN'16} &
                \textbf{TOR'16} &
                \textbf{TFC'16} &
                \textbf{CP'17} &
                \textbf{Other} &
                \textbf{Private} \\ 
            
            \midrule
            
                \cite{raw_info_first_study} &
                2015 &
                \CIRCLE  &
                \CIRCLE  &
                \CIRCLE  &
                \CIRCLE  &
                1000 &
                - &
                - &
                - &
                - &
                \textbf{T1} &
                - &
                - &
                - &
                - &
                - &
                \ding{51} \\

                \cite{datasets_ustc_tfc} &
                2017 &
                \LEFTcircle &
                \LEFTcircle &
                \CIRCLE  &
                \CIRCLE  &
                784 &
                - &
                - &
                - &
                - &
                \textbf{T1} &
                - &
                - &
                \ding{51} &
                - &
                - &
                - \\

                \cite{end_to_end} &
                2017 &
                \LEFTcircle &
                \LEFTcircle &
                \CIRCLE  &
                \CIRCLE  &
                784 &
                - &
                - &
                - &
                - &
                \textbf{T1} &
                \ding{51} &
                - &
                - &
                - &
                - &
                - \\

                \cite{clstm} &
                2018 &
                - &
                \LEFTcircle &
                \LEFTcircle  &
                \CIRCLE  &
                784 &
                3 \ding{68} &
                - &
                - &
                \textbf{T2} &
                - &
                \ding{51} &
                - &
                - &
                - &
                - &
                - \\

                \cite{automatic_multi_task_learning} &
                2018 &
                \CIRCLE  &
                \CIRCLE  &
                \CIRCLE  &
                \CIRCLE  &
                1024 &
                - &
                - &
                - &
                - &
                \textbf{T1} &
                \ding{51} &
                - &
                - &
                - &
                \cite{dataset_ctu_13} \ding{65} &
                - \\

                \cite{mobile_encrypted_traffic_classification} &
                2019 &
                \CIRCLE  &
                \CIRCLE  &
                \CIRCLE  &
                \CIRCLE  &
                784 &
                - &
                - &
                - &
                - &
                \textbf{T1} &
                - &
                - &
                - &
                - &
                - &
                \ding{51} \\

                \cite{transport_layer_payload} &
                2019 &
                - &
                - &
                - &
                \CIRCLE  &
                784 &
                - &
                - &
                - &
                - &
                \textbf{T1} &
                \ding{51} &
                - &
                - &
                - &
                - &
                - \\

                \cite{a_session_packets_based} &
                2019 &
                \LEFTcircle &
                \LEFTcircle &
                \CIRCLE  &
                \CIRCLE  &
                784 &
                - &
                - &
                - &
                - &
                \textbf{T1} &
                \ding{51} &
                - &
                - &
                - &
                - &
                - \\

                \cite{image_based_encrypted_traffic} &
                2020 &
                \CIRCLE  &
                \CIRCLE  &
                \CIRCLE  &
                \CIRCLE  &
                784 &
                - &
                - &
                - &
                - &
                \textbf{T1} &
                \ding{51} &
                - &
                - &
                - &
                - &
                - \\

                \cite{encrypted_network_traffic_classification_using} &
                2020 &
                - &
                \LEFTcircle &
                \CIRCLE  &
                \CIRCLE  &
                1500 &
                - &
                \textbf{T1} &
                - &
                - &
                - &
                \ding{51} &
                - &
                - &
                - &
                - &
                - \\

                \cite{transnet} &
                2020 &
                \LEFTcircle &
                \LEFTcircle &
                \CIRCLE  &
                \CIRCLE  &
                900 &
                - &
                - &
                - &
                - &
                \textbf{T1} &
                - &
                - &
                \ding{51} &
                - &
                - &
                - \\

                \cite{deep_packet} &
                2020 &
                - &
                \LEFTcircle &
                \CIRCLE  &
                \CIRCLE  &
                1500 &
                - &
                \textbf{T1} &
                - &
                - &
                - &
                \ding{51} &
                - &
                - &
                - &
                - &
                - \\

                \cite{foundation_pert} &
                2020 &
                \LEFTcircle &
                \LEFTcircle &
                \CIRCLE  &
                \CIRCLE  &
                128 &
                5 \ding{72} &
                - &
                - &
                \textbf{T3} &
                - &
                \ding{51} &
                - &
                - &
                - &
                - &
                \ding{51} \\

                \cite{dl_and_rl} &
                2020 &
                \CIRCLE &
                \CIRCLE &
                \CIRCLE  &
                \CIRCLE  &
                784 &
                - &
                - &
                - &
                - &
                \textbf{T1} &
                - &
                - &
                - &
                - &
                \cite{ctu_mixed_dataset} \ding{65} &
                \\

                \cite{cscnn} &
                2021 &
                - &
                \CIRCLE &
                \CIRCLE  &
                \CIRCLE  &
                1480 &
                - &
                \textbf{T1} &
                - &
                - &
                - &
                \ding{51} &
                - &
                - &
                - &
                - &
                - \\

                \cite{iclstm} &
                2021 &
                \LEFTcircle &
                \LEFTcircle &
                \CIRCLE  &
                \CIRCLE  &
                784 &
                - &
                - &
                - &
                - &
                \textbf{T1}  &
                \ding{51} &
                - &
                - &
                - &
                - &
                - \\

                \cite{cbd} &
                2021 &
                - &
                - &
                - &
                \CIRCLE &
                256 &
                10 \ding{68} &
                - &
                - &
                \textbf{T3} &
                - &
                \ding{51} &
                - &
                - &
                - &
                - &
                - \\

                \cite{deep_feature_based} &
                2021 &
                \CIRCLE &
                \CIRCLE &
                \CIRCLE &
                \CIRCLE &
                784 &
                - &
                - &
                - &
                \textbf{T1} &
                - &
                - &
                - &
                \ding{51} &
                - &
                \cite{ids_17} \ding{65} \cite{ids_12} \ding{65} &
                - \\

                \cite{global_aware_prototypical} &
                2022 &
                \LEFTcircle &
                \LEFTcircle &
                \CIRCLE  &
                \CIRCLE  &
                784 &
                - &
                - &
                - &
                \textbf{T1} &
                - &
                - &
                - &
                \ding{51} &
                - &
                - &
                - \\

                \cite{foundation_et_bert} &
                2022 &
                - &
                - &
                \LEFTcircle  &
                \CIRCLE  &
                128 &
                5 \ding{72} &
                \textbf{T1} &
                - &
                - &
                \textbf{T3} &
                \ding{51} &
                \ding{51} &
                \ding{51} &
                \ding{51} &
                \cite{foundation_et_bert} &
                - \\

                \cite{few_fine} &
                2022 &
                \CIRCLE &
                \CIRCLE &
                \CIRCLE  &
                \CIRCLE  &
                300 &
                - &
                - &
                - &
                - &
                \textbf{T1} &
                - &
                - &
                \ding{51} &
                - &
                - &
                - \\

                \cite{eetc} &
                2023 &
                \LEFTcircle &
                \LEFTcircle &
                \CIRCLE  &
                \CIRCLE  &
                3072 &
                - &
                - &
                - &
                - &
                \textbf{T1} &
                \ding{51} &
                - &
                - &
                - &
                - &
                \ding{51} \\

                \cite{rp_bert} &
                2023 &
                - &
                - &
                \LEFTcircle  &
                \CIRCLE  &
                128 &
                5 \ding{72} &
                \textbf{T1} &
                - &
                - &
                \textbf{T3} &
                - &
                - &
                - &
                - &
                \cite{datasets_bot_iot} \ding{65} &
                - \\
                
                \cite{bfcn} &
                2023 &
                - &
                - &
                \LEFTcircle  &
                \CIRCLE  &
                128 &
                - &
                \textbf{T1} &
                - &
                - &
                - &
                \ding{51} &
                - &
                - &
                - &
                - &
                - \\

                \cite{foundation_flow_mae} &
                2023 &
                \LEFTcircle &
                \LEFTcircle &
                \LEFTcircle  &
                \CIRCLE  &
                1024 &
                - &
                - &
                \textbf{T1} &
                - &
                - &
                \ding{51} &
                \ding{51} &
                \ding{51} &
                \ding{51} &
                \cite{datasets_cic_ids2018} \ding{65} &
                \\

                \cite{spatial_temporal} &
                2023 &
                - &
                \LEFTcircle &
                \CIRCLE  &
                \CIRCLE  &
                300 &
                30 \ding{68} &
                - &
                - &
                - &
                \textbf{T2} &
                - &
                - &
                - &
                - &
                \cite{cicandmal17} \ding{65} &
                \\
                
                \cite{foundation_yatc_full} &
                2024 &
                - &
                \LEFTcircle &
                \LEFTcircle  &
                \CIRCLE  &
                320 &
                5 \ding{72} &
                - &
                - &
                \textbf{T3} &
                - &
                \ding{51} &
                \ding{51} &
                \ding{51} &
                - &
                \cite{datasets_cic_iot2022} &
                - \\
                
            \bottomrule

            \addlinespace[4pt]
            \multicolumn{18}{l}{ 

                \begin{tabular}[l]{@{}l@{}}

                    \rowcolor{white}
                    
                    \large\textbf{Raw Features Used}: 
                    \textbf{L2}=Ethernet layer; 
                    \textbf{L3}=Network layer; 
                    \textbf{L4}=Transport layer; 
                    \textbf{L7}=Encrypted payload; 
                    \CIRCLE=Use with SII;
                    \LEFTcircle=Use without SII;
                    \\[5pt]
                    
                    \large\textbf{Data Extraction Strategy}: 
                    \textbf{\# Bytes}=No. of bytes; \hspace{0.2cm} 
                    \textbf{\# Packets}=No. of packets; \hspace{0.2cm} 
                    $n$ \ding{72}=First $n$ packets; \hspace{0.2cm} 
                    $n$ \ding{68}=Any consecutive $n$ packets;\\[3pt]
                    \hspace{4.8cm}                           \large \textbf{T1}=Type 1(First $m$ bytes); \hspace{0.2cm} 
                    \textbf{T2}=Type 2(First $m$ bytes of $n$ packets); \hspace{0.2cm} 
                    \textbf{T3}=Type 3(First $m$ bytes per packet of $n$ Packets); \hspace{0.2cm}     
                     \\[5pt]
                    
                    
                    \large\textbf{Datasets Used}: 
                    \textbf{VPN'16}: ISCXVPN2016 \cite{datasets_iscx_vpn}; 
                    \textbf{Tor'16}: ISCXTor2016 \cite{datasets_iscx_tor}; 
                    \textbf{TFC'16}: USTC-TFC2016 \cite{datasets_ustc_tfc}; 
                    \textbf{CP'17}: Cross-Platform Application \cite{datasets_cross_platform};  \\[3pt]
                    \hspace{2.7cm}  \large\textbf{\ding{65}}: Dataset collected on or before 2018
                    
                \end{tabular}
                
            }
            
        \end{tabular}
    }
    
\end{table*}

\section{Unresolved Challenges in NTC}
\label{sec:unresolved_challenges}
Despite significant advances, several open problems inhibit the development of robust and generalizable classifiers. These issues impede the ability of models to perform effectively in real-world scenarios and limit the practical applicability of research findings. This section identifies and critically examines three primary snags affecting current NTC approaches: reliance on outdated datasets, unsubstantiated assumptions, and issues caused by oversights in design choices.

\subsection{Snag 1: Legacy Datasets}
\label{subsec:snag_2_outdated_datasets}
The literature summarized in Table \ref{tab:1_lit_summary} reveals that many studies continue to rely on network traffic datasets collected before 2018. This reliance raises concerns since encryption protocols and cipher suites have evolved significantly since then. Specifically, the adoption of protocols such as TLS 1.3 has deprecated algorithms such as 3DES and RC4 and has mandated or strongly recommended the use of enhanced cipher suites. Furthermore, TLS 1.3 encrypts portions of the handshake process previously transmitted in plaintext \cite{tls_13}. 

These changes reduce the metadata and observable patterns available to network traffic classifiers. Consequently, models developed using older datasets and the assumptions they are based on, may not adequately account for these changes. Applying these models to contemporary network traffic could lead to decreased performance or even complete ineffectiveness.

To address this potential gap, we evaluate the validity and applicability of older datasets in the context of current network conditions. To this end, we formulate the following research questions to determine whether newer datasets are necessary for improving network traffic classification.

\begin{tcolorbox}[colback=gray!5, colframe=gray!40, title=\textbf{S1 - Research Questions}, fonttitle=\bfseries\color{black}]
\textbf{S1-RQ1:} Do legacy datasets contain encrypted network traffic that accurately reflects modern protocols? \vspace{0.2cm} \\ 
\textbf{S1-RQ2:} Do encryption algorithms used in legacy datasets remain valid and relevant for contemporary NTC?
\end{tcolorbox}

\subsection{Snag 2: Oversights in Design Choices}
\label{subsec:snag_3_design_choices}
We suggest that NTC models often overfit due to a lack of careful feature selection. Feature choices are often made without considering the selected traffic granularity and data extraction strategy. We categorize this tendency to overfit into three types: data leakage overfitting, contextual overfitting, and temporal overfitting.

\subsubsection{Data Leakage Overfitting}
\label{subsubsec:data_leakage_overfitting}
This problem occurs when models inadvertently learn from features that should not be available during inference, which leads to an illusion of high performance that does not generalize to real-world scenarios. While some studies have made commendable efforts to prevent overfitting related to Strong Identification Information (SII) (see Table \ref{tab:1_lit_summary}), some have not. Another issue arises from exposing Server Name Indication (SNI) \footnote{The SNI extension in the TLS handshake reveals the hostname the client is attempting to connect to} to ML models. As SNI is frequently used for labelling traffic in datasets \cite{foundation_et_bert, dataset_sni_based_labelling_1, dataset_sni_based_labelling_2}, exposing it creates a potential shortcut that models can exploit.

Studies that perform classification based on the \textit{first $m$ bytes} of a \textit{flow} or \textit{session} granularity, where $m$ is set to include the initial TCP and TLS handshakes frequently do not obfuscate the SNI. To capture the SNI, $m$ must be large enough to encompass the cumulative sizes of the packets from the beginning of the TCP session up to and including the TLS Client Hello message. The TCP handshake involves three packets (SYN, SYN-ACK, ACK) with minimal payloads, typically totalling around 162 bytes (without TCP options) \cite{tcp_rfc9293}. The TLS Client Hello message, which includes the SNI, follows the TCP handshake and varies in size but is typically around 600 bytes, depending on the number of supported cipher suites and extensions \cite{tls_13, tls_12}. Therefore, choosing an $m$ value greater than 700 bytes generally ensures that the SNI is included in the training data. According to Table \ref{tab:1_lit_summary}, $m$ ranges from 764 to 3072 in literature, raising concerns about data leakage overfitting.

Similarly, studies that consider the \textit{first $n$ packets} of a \textit{flow} or \textit{session} granularity, where $n$ includes the packets containing the TLS Client Hello message, are likely to capture the SNI. Since the TCP handshake consists of three packets and the TLS Client Hello is typically sent in the fourth packet, setting $n \geq 4$ is likely to include the SNI \cite{tls_13}. 

If the SNI is not obfuscated, models may end up using SNI values as shortcuts for classification rather than learning meaningful patterns in the underlying traffic. However, in practice, the presence of the TLS handshake and the SNI is not always guaranteed. For example, in TLS 1.3, mechanisms such as Encrypted Client Hello (ECH) encrypt the SNI to enhance privacy \cite{tls_esni}. Further, the presence of TLS handshake is not always guaranteed due to session reestablishment \cite{dataset_sni_based_labelling_1}. 

Consequently, models that rely on the SNI may struggle with accurate classification in real-world conditions where this information is either encrypted or absent.

\subsubsection{Contextual Overfitting}
\label{subsubsec:contextual_overfitting}
Contextual overfitting refers to the phenomenon where a model learns to exploit irrelevant features or context-dependent patterns present in the data that do not reflect the intrinsic nature of the target class. In the NTC domain, contextual overfitting arises when models exploit features that are artifacts of the network protocols or specific implementations rather than intrinsic properties of the traffic generated by target applications.

In contrast to data leakage overfitting, contextual overfitting impacts classifiers of studies that split a single TCP/UDP session into multiple training and testing samples. As shown in Table \ref{tab:1_lit_summary}, they are (1) \textit{packet} and \textit{burst} granularity-based classifiers, and (2) \textit{flow} and \textit{session} granularity-based classifiers which extract information from \textit{any consecutive $n$ packets}.

For example, according to RFC 791 \cite{ip_rfc791}, RFC 6864 \cite{ip_rfc6864} and RFC 6274 \cite{ip_rfc6274}, the IP Identification (IP ID) field in \textit{L3} (IP header), is initialized for each session with a pseudo-random number generator and incremented sequentially for each packet transmitted. As a consequence, the high-order bits remain relatively consistent.

Similarly, the IP Header Checksum, which is a 16-bit one's complement of the one's complement sum of all 16-bit words in \textit{L3} (IP header) \cite{ip_rfc791}, poses a risk for contextual overfitting. Due to the changing nature of attributes, such as the IP ID and Total Length, the Header Checksum varies with each packet. However, packets with similar Total Lengths and minor differences in IP IDs (i.e., adjacent packets) have similar checksum values within a session. 


As per RFC 9293 \cite{tcp_rfc9293}, fields such as the Sequence Number and Acknowledgment Number in \textit{L4} (TCP header) are specific to each TCP session, initialized using pseudo-random functions and incremented with data transfer. Due to gradual increments caused by uploads/downloads, the high-order digits of Sequence and Acknowledgement numbers remain mostly consistent across packets in the same TCP session.


As a consequence, models that see multiple packets from the same session may inadvertently learn patterns based on the stable high-order bits of fields like IP ID, IP header checksum, and TCP Sequence and Acknowledgment numbers, leading to overfitting. Since these features are either randomly generated per session or influenced by such random values, they should not be relied upon for classification, as they reflect session-specific artifacts rather than meaningful traffic characteristics.

\subsubsection{Temporal Overfitting}
\label{subsubsec:temporal_overfitting}
Temporal overfitting occurs when models capture features that fluctuate over time due to dynamic network conditions, system configurations, or temporal changes unrelated to the application's behaviour. These features may provide superficial patterns specific to the time and environment of data collection, leading to models that do not generalize well under different conditions.

A notable example is the TCP Timestamp Option, introduced in RFC 1323 \cite{tcp_rfc1323} to enhance TCP performance over high-speed networks. Each TCP segment can contain a 32-bit Timestamp Value (TSval) set by the sender and a 32-bit Timestamp Echo Reply (TSecr) from the remote host. The TSecr field reflects the TSval received from the sender in the previous segment, allowing for more accurate round-trip time (RTT) measurements. Since packets within a session are sent in quick succession, the most significant bits of TSval and TSecr remain consistent.

Further, TCP Window Size is a dynamic value negotiated between the sender and receiver to manage flow control and optimize data transmission. According to RFC 7323 \cite{tcp_rfc7323} and RFC 9438 \cite{tcp_rfc9438}, the Window Size is influenced by several factors independent of the traffic class, such as the available receive buffer, network congestion levels, bandwidth-delay product (BDP), and the Path Maximum Transmission Unit (MTU). For instance, during periods of high network congestion, the Window Size may be reduced to prevent packet loss. If models rely on the TCP Window Size as a feature, they could overfit to transient network conditions upon which the dataset is collected.

Since these features fluctuate based on time-sensitive factors such as timing within a session, network congestion, and buffer availability, they should not be relied upon for classification. These elements reflect temporary network states rather than intrinsic characteristics of the traffic itself, thus limiting the model’s ability to generalize across varying network environments. \\

Table \ref{tab:4_overfitting_vs_granularity} summarizes the overfitting tendencies associated with the discussed raw features. It details the types of overfitting each feature may introduce, along with the affected traffic granularities and data extraction strategies. Notably, none of the studies listed in Table \ref{tab:1_lit_summary} take steps to obfuscate or mask these session-specific attributes. As a consequence, while these features may contribute to higher performance on familiar data, models trained on them risk failing to generalize to unseen traffic where these learned patterns are absent. To validate our suppositions, we formulate the following research questions.

\begin{table}[t]
\renewcommand{\arraystretch}{1.3}

    \caption{Overfitting Type and affected traffic granularity}
    \label{tab:4_overfitting_vs_granularity}
    
    \centering

    \resizebox{\columnwidth}{!}{
        \rowcolors{2}{}{lightgray}        
        
        \begin{tabular}{*{1}{L{3.2cm}}*{1}{C{0.8cm}}*{2}{C{1cm}}*{2}{C{2.1cm}}} 
        
            \toprule
                
                \multicolumn{1}{c}{\multirow{2}{*}{\textbf{Feature}}} & 
                \multirow{2}{*}{\textbf{Type}} & 
                \multicolumn{4}{c}{\textbf{Affected Granularity}} \\ 

                \cmidrule(lr){3-6}

                \rowcolor{white}
                &
                &
                \textbf{Packet} &
                \textbf{Burst} &
                \textbf{Flow} &
                \textbf{Session} \\

            \midrule

                Src. MAC Addr. & 
                DL& 
                \ding{51} &
                \ding{51} &
                \ding{51} &
                \ding{51} \\

                Dst. MAC Addr. & 
                DL& 
                \ding{51} &
                \ding{51} &
                \ding{51} &
                \ding{51} \\

                IP ID & 
                C& 
                \ding{51} &
                \ding{51} &
                \textbf{T2}\ding{68}  \textbf{T3}\ding{68}&
                \textbf{T2}\ding{68}  \textbf{T3}\ding{68} \\

                IP Header Checksum & 
                C& 
                \ding{51} &
                \ding{51} &
                \textbf{T2}\ding{68}  \textbf{T3}\ding{68} &
                \textbf{T2}\ding{68}  \textbf{T3}\ding{68} \\

                Src. IP Addr. & 
                DL& 
                \ding{51} &
                \ding{51} &
                \ding{51} &
                \ding{51} \\

                Dst. IP Addr. & 
                DL& 
                \ding{51} &
                \ding{51} &
                \ding{51} &
                \ding{51} \\

                Src. Port & 
                DL& 
                \ding{51} &
                \ding{51} &
                \ding{51} &
                \ding{51} \\

                Dst. Port & 
                DL& 
                \ding{51} &
                \ding{51} &
                \ding{51} &
                \ding{51} \\

                TCP Seq. No. &
                C& 
                \ding{51} &
                \ding{51} &
                \textbf{T2}\ding{68}  \textbf{T3}\ding{68} &
                \textbf{T2}\ding{68}  \textbf{T3}\ding{68} \\

                TCP Ack. No. &
                C& 
                \ding{51} &
                \ding{51} &
                \textbf{T2}\ding{68}  \textbf{T3}\ding{68} &
                \textbf{T2}\ding{68}  \textbf{T3}\ding{68} \\
                
                TCP Window Size & 
                T& 
                \ding{51} &
                \ding{51} &
                \textbf{T2}\ding{68}  \textbf{T3}\ding{68} &
                \textbf{T2}\ding{68}  \textbf{T3}\ding{68} \\

                TCP Options - TSval & 
                T& 
                \ding{51} &
                \ding{51} &
                \textbf{T2}\ding{68}  \textbf{T3}\ding{68} &
                \textbf{T2}\ding{68}  \textbf{T3}\ding{68} \\

                TCP Options - TSerc & 
                T& 
                \ding{51} &
                \ding{51} &
                \textbf{T2}\ding{68}  \textbf{T3}\ding{68} &
                \textbf{T2}\ding{68}  \textbf{T3}\ding{68} \\
                
                TLS SNI & 
                DL& 
                - &
                - &
                \textbf{T1} \textbf{T2}\ding{72} \textbf{T3}\ding{72}&
                \textbf{T1} \textbf{T2}\ding{72} \textbf{T3}\ding{72}\\
            
            \bottomrule

            \addlinespace[4pt]
            \multicolumn{6}{l}{ 

                \begin{tabular}[l]{@{}l@{}}

                    \rowcolor{white}
                    \large \textbf{DL}=Data Leakage; \hspace{0.2cm} 
                    \textbf{C}=Contextual; \hspace{0.2cm} 
                    \textbf{T}=Temporal; \\[3pt]

                    \large \ding{72}=First $n$ Packets; \hspace{0.2cm} 
                    \ding{68}=Any consecutive $n$ Packets;\\[3pt]
                    
                    \large \textbf{T1}=Type 1(First $m$ bytes); \hspace{0.2cm} 
                    \textbf{T2}=Type 2(First $m$ bytes of $n$ packets); \hspace{0.2cm} \\[3pt]
                    
                    \large \textbf{T3}=Type 3(First $m$ bytes per packet of $n$ Packets); \hspace{0.2cm} \\[3pt]

                    \large \ding{51}=Affects regardless of the data extraction strategy.
                    
                \end{tabular}
                
            }
        
        \end{tabular}
    }
\end{table}

\begin{tcolorbox}[colback=gray!5, colframe=gray!40, title=\textbf{S2 - Research Questions}, fonttitle=\bfseries\color{black}]
    \textbf{S2-RQ1:} Does data leakage from SII and SNI affect the generalizability and robustness of network traffic classifiers? \vspace{0.2cm} \\
    \textbf{S2-RQ2:} Do session-specific contextual artifacts contribute to overfitting in NTC? \vspace{0.2cm} \\
    \textbf{S2-RQ3:} Do time-specific temporal artifacts contribute to overfitting in NTC?
\end{tcolorbox}

\subsection{Snag 3: Unsubstantiated Assumptions}
\label{subsec:snag_1_unsubstantiated_assumptions}
Examining the literature reveals several unsubstantiated and often contradictory assumptions regarding design choices.

While all studies shown in Table \ref{tab:1_lit_summary} utilize the encrypted payload (\textit{L7}), their justifications vary. 

Some studies suggest that encrypted payloads contain inherent patterns resulting from the imperfect randomness of encryption algorithms \cite{encrypted_network_traffic_classification_using, deep_packet, cbd, foundation_et_bert, bfcn, foundation_flow_mae, foundation_yatc_full}. 

However, TLS 1.3 guarantees that the same plaintext will always produce different ciphertexts due to the use of Authenticated Encryption with Associated Data (AEAD) ciphers like AES-128-GCM, AES-256-GCM and ChaCha20-Poly1305, and unique initialization vectors (IVs) \cite{tls_13}. Given these robust security measures, the claim that machine learning models can still learn and exploit patterns directly from encrypted payloads is concerning. It suggests potential vulnerabilities in the cipher suites, implying that encrypted communications may not be as secure or random as intended.

There is also inconsistency in the perceived sufficiency of encrypted payloads for classification: several studies argue that both metadata (\textit{L2}, \textit{L3}, and \textit{L4}) and encrypted payloads (\textit{L7}) are necessary for accurate classification. In contrast, others assert that the encrypted payload alone is adequate \cite{cbd, transport_layer_payload}.

Furthermore, all studies listed in Table \ref{tab:1_lit_summary} use truncation and padding techniques to create a consistent, fixed-length embedding for training proposed ML models. However, some studies argue that altering payload lengths to fit a fixed-size representation can result in losing important information, negatively impacting the model's classification performance \cite{markov-gan}.

To address these discrepancies, we formulate the following research questions:
\begin{tcolorbox}[colback=gray!5, colframe=gray!40, title=\textbf{S3 - Research Questions}, fonttitle=\bfseries\color{black}]
    \textbf{S3-RQ1:} Can state-of-the-art network traffic classifiers detect meaningful patterns in encrypted payloads? \vspace{0.2cm} \\ 
    \textbf{S3-RQ2:} Is encrypted payload alone sufficient for accurate network traffic classification? \vspace{0.2cm} \\
    \textbf{S3-RQ3:} Does truncating or padding traffic payloads impact classification accuracy?
\end{tcolorbox}

\section{Decoding the NTC Fidelity}
\label{sec:decoding_the_ntc_fidelity}
In the previous section, we identified research questions related to: (\textit{Snag-1}) limitations of widely used datasets, (\textit{Snag-2}) oversights in design choices, and (\textit{Snag-3}) contradicting assumptions in literature. To address the \textit{Snag-1} questions, we begin with an evaluation of widely used datasets, assessing their relevance and content in light of these concerns. Following this, we conduct 348 strategic experiments using state-of-the-art classifiers to answer the questions posed in \textit{Snag-2} and \textit{Snag-3}. 

\subsection{NTC Dataset Evaluation}
In Section \ref{subsec:snag_2_outdated_datasets}, we raised concerns about the credibility and applicability of network traffic classification models that rely on datasets collected before 2018. This section addresses these doubts by empirically evaluating datasets to uncover their true potential and suitability for contemporary NTC tasks.

\subsubsection{Dataset Selection}
To conduct meaningful analysis, we select public datasets used in more than one study presented in Table \ref{tab:1_lit_summary} or those claimed to include TLS 1.3 traffic. This selection criterion ensures that we focus on datasets with significant influence in the research community and those containing modern encryption protocols. By scrutinizing these datasets, we aim to answer research questions \textit{S1-RQ1} and \textit{S1-RQ2}, thereby justifying their use in developing effective NTC models.

\subsubsection{Methodology}
To systematically evaluate the encryption protocols and encryption algorithms present in the selected datasets, we process packet capture (PCAP) files and extract relevant encryption information using the Tshark tool\footnote{Tshark: https://tshark.dev/}. A simplified pseudocode of our analysis script is presented in Algorithm \ref{alg:1_dataset_evaluation}.

We extract all unique session identifiers for each PCAP file in the dataset, which represent individual communication sessions in the network traffic \textit{(Line 2, 3)}. These session IDs encompass TCP and UDP sessions, allowing for a comprehensive dataset analysis.

We then iterate over each session ID to determine whether the session is encrypted \textit{(Line 4, 5)}. This task is carried out by checking for the presence of encryption protocols such as TLS for TCP sessions or DTLS/QUIC for UDP sessions \textit{(Line 6)}.

Encrypted sessions: If a session is encrypted (\textit{type} is \textit{encrypted}), we increment the count of encrypted sessions of the dataset being analyzed. We then attempt to identify the specific cipher suite used in the session by examining the "Server Hello" packet of the TLS handshake packets \textit{(Line 7, 8)}. Suppose the cipher suite cannot be identified due to missing handshake packets. In that case, we increment the count of sessions with \textit{unknown} to account for sessions with unidentified encryption methods \textit{(Line 9, 10)}. If the cipher suite can be determined, we extract the encryption algorithm and increment the count of sessions encrypted by it \textit{(Line 12, 13)}. This helps us understand the distribution and representation of different encryption algorithms within a dataset.

Unencrypted sessions: If a session is not encrypted (\textit{type} is \textit{unencrypted}), we increment the count of \textit{unencrypted} sessions. This category includes sessions either transmitted in plaintext or unrelated to encryption protocols \textit{(Line 16)}.

After processing all sessions within a PCAP file, we continue to the next file in the dataset \textit{(Line 18)}. By aggregating the statistics across all PCAP files, we construct a detailed overview of encryption usage and cipher suite distribution within the dataset.

\begin{algorithm}[tb]
    \caption{Dataset evaluation} 
    \label{alg:1_dataset_evaluation}
    {
        \setstretch{1.1}
        \begin{algorithmic}[1]
            \State Initialize \textit{pcap\_stats} dictionary for the dataset
            \For{each \textit{pcap\_file} in dataset}
                \State \textit{session\_ids} $\leftarrow$ Extract session IDs from \textit{pcap\_file}
                \For{each \textit{s\_id} in \textit{session\_ids}}
                    \State \textit{type} $\leftarrow$ Check if session \textit{s\_id} is encrypted
                    \If{\textit{type} is "encrypted"}
                        \State Increment \textit{pcap\_stats["encrypted"]} by $1$
                        \State \textit{c\_suite} $\leftarrow$ Get cipher suite of session \textit{s\_id}
                        \If{\textit{c\_suite} is "unknown"}
                            \State Increment \textit{pcap\_stats["unknown"]} by $1$
                        \Else
                            \State \textit{e\_algo} $\leftarrow$ Get enc. algo from \textit{c\_suite}
                            \State Increment \textit{pcap\_stats[\textit{e\_algo}]} by $1$
                        \EndIf
                    \Else
                        \State Increment \textit{pcap\_stats["unencrypted"]} by $1$
                    \EndIf
                \EndFor
            \EndFor
        \end{algorithmic}
    }
\end{algorithm}

\subsubsection{\textit{S1-RQ1}: Encryption in Legacy Datasets}
\begin{figure*}[!t]
    \centering
    
    \subfloat[Encryption usage]{\includegraphics[width=0.95\columnwidth]{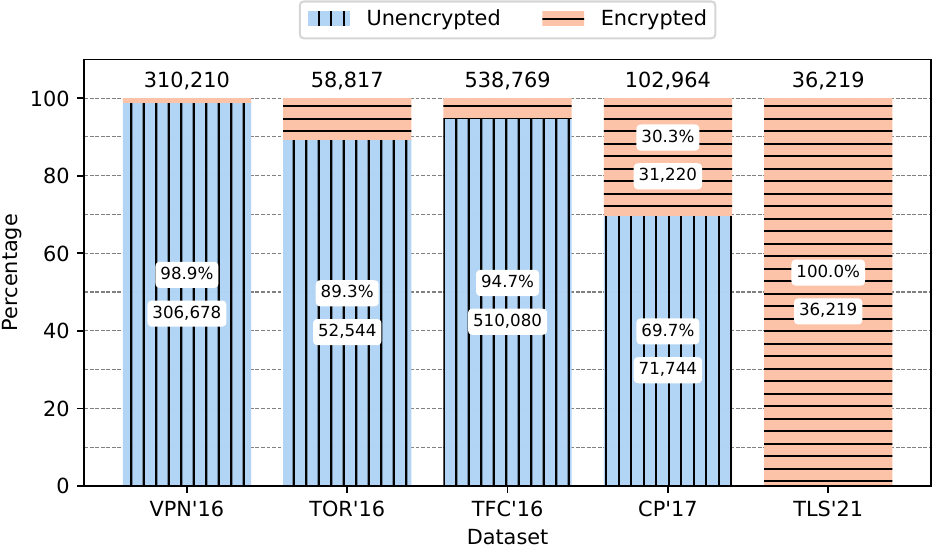}%
    \label{img:5_dataset_analysis_encryption_usage}}
    \hfil
    \hspace{1.2em}
    \subfloat[Cipher suite distribution]{\includegraphics[width=\columnwidth]{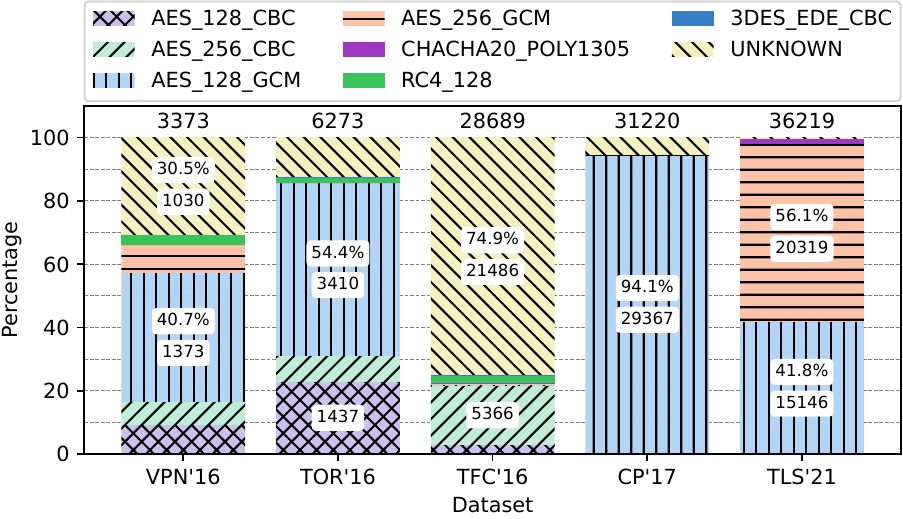}%
    \label{img:5_dataset_analysis_cipher_suite_distribution}}

    \vspace{1.5em}
    \noindent\footnotesize\textbf{VPN'16}: ISCXVPN2016; \textbf{Tor'16}: ISCXTor2016; \textbf{TFC'16}: USTC-TFC2016; \textbf{CP'17}: Cross-Platform Application; \textbf{TLS'21}: CSTNET-TLS1.3; 

    \vspace{0.5em}
    \justifying\noindent Each segment of the stacked bars represents a percentage of sessions. The total number of sessions relevant for the analysis is displayed on top of the corresponding stacked bar. Additionally, segments indicating the number of sessions are labeled if they constitute more than 15\% of the total.
    
    \caption{Public network traffic dataset evaluation.}
    \label{5_dataset_analysis}
\end{figure*}

The results of the encryption usage analysis of the selected datasets are shown in Figure \ref{img:5_dataset_analysis_encryption_usage}. This indicates a substantial variation in the proportion of unencrypted versus encrypted traffic across different datasets. The results underscore the prevalence of unencrypted traffic in several widely used datasets, undermining their applicability in evaluating modern encrypted NTC models.

Among the datasets analyzed, the CSTNET-TLS1.3 dataset \cite{foundation_et_bert} stands out as it exclusively contains encrypted network traffic, as claimed.

In stark contrast, several legacy datasets reveal a significant presence of unencrypted traffic. The ISCXVPN2016 \cite{datasets_iscx_vpn}, and USTC-TFC2016 \cite{datasets_ustc_tfc} datasets, for example, exhibit notably high percentages of unencrypted traffic, at 98.9\% and 94.7\%, respectively. This suggests a dominant presence of plaintext communications in these datasets, which limits their applicability in scenarios that require encrypted traffic analysis. Similarly, the ISCXTor2016 \cite{datasets_iscx_tor} dataset, which is intended to capture traffic from the Tor network, still shows 89.3\% of its traffic as unencrypted. This indicates that a considerable portion of the dataset lacks the encryption expected from a privacy-focused network like Tor. The Cross-Platform Application dataset \cite{datasets_cross_platform}, which is relatively more balanced, still reveals that 69.7\% of its traffic remains unencrypted. 

These proportions are widely misaligned with real-world conditions. According to Google's Transparency Report, as of October 2024, more than 93\% of web pages loaded in Google Chrome are secured with SSL/TLS encryption \cite{google_report}. This discrepancy highlights a critical gap between the encryption distributions within these datasets and the actual state of internet traffic, where encryption has become the norm.

\subsubsection{\textit{S1-RQ2}: Encryption Algorithm Usage}
As shown in Figure \ref{img:5_dataset_analysis_cipher_suite_distribution}, CSTNET-TLS1.3 dataset stands out by exclusively containing traffic encrypted with current, secure AEADs, in line with the recommendations of TLS 1.3 \cite{tls_13}. In contrast, pre-2018 datasets like ISCXVPN2016, ISCXTor2016, and USTC-TFC2016 contain traffic sessions encrypted using outdated algorithms such as AES with CBC mode (both 128-bit and 256-bit), 3DES, and RC4, which are deprecated in contemporary security standards due to their known vulnerabilities \cite{rc4_deprication, 3des_deprication, tls10_tls11_deprecation}. This reliance on outdated encryption algorithms further limits the effectiveness of legacy datasets for developing and benchmarking classifiers.

Moreover, another notable gap is the limited representation of traffic encrypted with the ChaCha20-Poly1305 cipher suite \cite{chacha20}. Despite being a recommended algorithm in TLS 1.3 and the default AEAD in several widely-used secure communication protocols like OpenSSH, WireGuard, and OTRv4, none of the datasets include considerable proportions of ChaCha20-Poly1305-encrypted traffic. This lack of representation is a critical omission, considering the growing adoption of this algorithm in modern security protocols \cite{chacha20_gaining_traction, tls_13}.

\textit{\textbf{Takeaway 1}: Substantial portions of public datasets contain unencrypted traffic.}

\textit{\textbf{Takeaway 2}: Widely-used datasets include sessions encrypted by vulnerable and deprecated ciphers, potentially misleading machine learning models.}

\textit{\textbf{Takeaway 3}: Although secure ciphers like ChaCha20-Poly1305 are seeing growing adoption, they remain underrepresented in public datasets.}

\textit{\textbf{Takeaway 4}: Given the widespread adoption of TLS 1.3 and QUIC \cite{tls_13_adaptation, quic_getting_traction}, public datasets should be critically evaluated and updated to reflect contemporary conditions.}

\subsection{CipherSpectrum}
The analysis in the previous sections highlighted several limitations in existing public datasets, particularly their unencrypted nature and usage of outdated encryption protocols. Moreover, the literature indicates that many publicly available datasets are flow features (i.e., extracted statistics) instead of raw traffic \cite{datasets_ustc_tfc, cscnn}. While these datasets are valuable for statistics-based NTC, they are unsuitable for raw information-based or multimodal methods. As a result, many researchers are compelled to rely on self-collected or private traffic datasets, which compromises the reproducibility and credibility of their results \cite{cscnn, end_to_end, clstm, a_session_packets_based}. Further, some publicly available datasets, such as ISCXVPN2016, exhibit significant class imbalances \cite{deep_packet}, which can be detrimental to the performance of deep-learning models that are known to be sensitive to such imbalances \cite{bfcn}.

To collectively address these issues and to support our subsequent experiments, we develop a new traffic dataset: \textit{CipherSpectrum}. CipherSpectrum includes network traffic encrypted with modern cipher suites mandated/strongly-recommended by TLS 1.3 \cite{tls_13}, providing a robust foundation for research in raw information-based NTC.

\subsubsection{Composition}
CipherSpectrum consists of encrypted TCP/UDP sessions for 40 distinct domains (classes), each represented by traffic encrypted with the three major TLS 1.3 cipher suites. Specifically, for each class and cipher suite, CipherSpectrum contains 1,000 TCP/UDP session samples, totalling 120,000 sessions (40 classes $\times$ 3 cipher suites per class $\times$ 1,000 sessions per suite). For instance, traffic for \textit{example.com} includes 1,000 sessions with TLS-AES-128-GCM-SHA256, 1,000 with TLS-AES-256-GCM-SHA384, and 1,000 with TLS-CHACHA20-POLY1305-SHA256. To support the development of robust and generalizable NTC models and facilitate research that reflects a range of encrypted traffic scenarios, we make CipherSpectrum publicly available\footnote{CipherSpectrum: https://cspectrum.web.cse.unsw.edu.au}.

The data collection and development methodology for CipherSpectrum is provided in Appendix \ref{app_sec:cipherspectrum} to maintain focus on the primary discussion. 

\textit{\textbf{Takeaway 5}: CipherSpectrum provides a comprehensive dataset by uniformly representing traffic encrypted using the three mandated and recommended cipher suites of TLS 1.3.}

\textit{\textbf{Takeaway 6}: Public availability of CipherSpectrum supports the development of robust and generalizable NTC models, facilitating cipher-agnostic network traffic classification.}

\subsection{Design Choices and Assumptions Validation}
\label{sec:evaluation_2_design_choices_and_assumptions}
In this section, we address the research questions related to (\textit{Snag-2}) oversights in design choices and (\textit{Snag-3}) unsubstantiated assumptions, as discussed in Section \ref{subsec:snag_3_design_choices} and Section \ref{subsec:snag_1_unsubstantiated_assumptions} respectively.

We conduct 348 systematic occlusion experiments to substantiate the speculative claims in which the features in question are masked, removed, or obfuscated from the input. This approach allows us to assess the impact of these features on model performance and to prove our conjectures empirically.

\subsubsection{Dataset Selection}
For our evaluation, we focus on encrypted traffic adhering to TLS 1.3 by utilizing two datasets: CipherSpectrum and CSTNET-TLS1.3 \cite{foundation_et_bert}. Although both models demonstrate their multi-class classification abilities, we randomly select 10 classes from each dataset. This decision is motivated by several factors: (1) preliminary tests on larger class sets confirmed the same pitfalls, reinforcing the generalizability of our findings; (2) the pitfalls we highlight are independent of the class count, making a more extensive selection unnecessary; (3) restricting the class set aligns with YaTC’s proposed capabilities, which max out at 20 classes, ensuring a fair and unbiased comparison; and (4) conducting 348 occlusion experiments is computationally intensive, and limiting the number of classes allows for a thorough yet efficient analysis, enabling deeper scrutiny and reliable reproducibility. 

In CipherSpectrum, the selected classes include an equal mix of traffic encrypted using AES-128-GCM, AES-256-GCM, and CHACHA20-POLY1305, ensuring a balanced representation of encryption algorithms. In CSTNET-TLS1.3, we address the class imbalance problem by randomly selecting 10 classes containing more than 400 TCP/UDP sessions. Furthermore, we randomly select 400 TCP/UDP sessions from each selected class. By adopting a dual-dataset approach, we aim to confirm the generalizability of our findings across different traffic characteristics and encryption schemes. 

\subsubsection{Model Selection}
\label{subsubsec:model_selection}
To substantiate our claims, we selected ET-BERT \cite{foundation_et_bert} and YaTC \cite{foundation_yatc} for our evaluation due to their open-source nature, credibility from publication in top-ranked venues (A* by CORE), and proven superiority over traditional ML methods \cite{netbench, yatc_etbert_best_1, foundation_yatc}. ET-BERT, regarded as state-of-the-art \cite{foundation_flow_mae, foundation_yatc}, serves as a foundation for numerous studies \cite{bfcn, rp_bert, et_bert_rep_1, et_bert_rep_2, et_bert_rep_3}, while both models effectively address a range of NTC challenges. However, our goal is not to evaluate or compare their performance as an end objective. Instead, we use these models as illustrative tools to demonstrate how design choices (see Section \ref{subsec:preliminries_design_choices}) influence model behaviour, leading to overfitting and other methodological pitfalls. Appendix \ref{app_sec:data_preprocessing} explains the model-specific preprocessing steps applied to represent each level of granularity and data extraction strategy for ET-BERT and YaTC.

\subsubsection{Occlusion Strategies}
The occlusion strategies used in our analysis are summarised in Table \ref{tab:6_occlusion_strategies}. The \textit{A1} occlusion, which includes all data, provides a baseline performance metric. The \textit{D1} occlusion, designed to test Strong Identification Information (SII) based data leakage overfitting (see Section \ref{subsubsec:data_leakage_overfitting}), provides a secondary baseline. Our observations indicate that the presence of SII can overshadow other features, preventing models from learning additional information. Therefore, subsequent occlusions are compared against the performance of \textit{D1}. In \textit{D2}, we eliminate Server Name Indication (SNI) by replacing relevant bytes with random hex values, allowing us to examine the extent of SNI-based data leakage overfitting.


\begin{table*}[!t]
\renewcommand{\arraystretch}{1.4}

    \caption{Feature occlusion strategies}
    \label{tab:6_occlusion_strategies}
    
    \centering
    \small

    \resizebox{\textwidth}{!}{        
        \begin{tabular}{cl*{1}{C{1.3cm}}*{2}{C{1cm}}*{1}{C{1.3cm}}*{1}{C{0.7cm}}*{1}{C{1.6cm}}*{1}{C{1.5cm}}*{1}{C{1cm}}*{1}{C{1.1cm}}*{1}{C{.5cm}}*{1}{L{4.5cm}}}

            \toprule

                \multirow{2}{*}{\textbf{ID}} &
                \multirow{2}{*}{\textbf{Occlusion Strategy}} &
                \multicolumn{1}{c}{\textbf{L2}} &
                \multicolumn{3}{c}{\textbf{L3}} &
                \multicolumn{4}{c}{\textbf{L4}} &
                \multicolumn{2}{c}{\textbf{L7}} & 
                \multirow{2}{*}{\textbf{Implications of the results}} \\

                \cmidrule(lr){3-3} 
                \cmidrule(lr){4-6} 
                \cmidrule(lr){7-10}
                \cmidrule(lr){11-12}

                \rowcolor{white}
                &
                &
                
                \textbf{MAC A.}&
                
                \textbf{IP A.} &
                \textbf{IP ID} &
                \textbf{Checksum} &

                \textbf{Ports} &
                \textbf{Seq \& Ack} &
                \textbf{Window S.} &
                \textbf{Options} &

                \textbf{Payload} &
                \textbf{SNI} &
                \\
            
            \midrule

                \rowcolor{lightgray}
                $A1$ &
                All Data &
                - &
                - &
                - &
                - &
                - &
                - &
                - &
                - &
                - &
                - &
                Baseline performance \\

                \cmidrule(lr){1-13}
                
                $D1$ &
                Anonymized SII &
                R &
                R &
                - &
                - &
                R &
                - &
                - &
                - &
                - &
                - &
                Reliance on SII \\

                \rowcolor{lightgray}
                $D2$ &
                Anonymized SNI &
                R &
                R &
                - &
                - &
                R &
                - &
                - &
                - &
                - &
                R &
                Reliance on SNI \\

                \cmidrule(lr){1-13}

                $C$ &
                w/o Contextual O. &
                R &
                R &
                R &
                R &
                R &
                R &
                - &
                - &
                - &
                - &
                Contextual overfitting proof \\

                \rowcolor{lightgray}
                $T$ &
                w/o Temporal O. &
                R &
                R &
                - &
                - &
                R &
                - &
                R &
                R &
                - &
                - &
                Temporal overfitting proof \\

                $CTD$ &
                w/o Overfitting &
                R &
                R &
                R &
                R &
                R &
                R &
                R &
                R &
                - &
                R &
                Absolute performance \\
                
                \cmidrule(lr){1-13}

                \rowcolor{lightgray}
                $H1$ &
                Header Only &
                R &
                R &
                - &
                - &
                R &
                - &
                - &
                - &
                E &
                R &
                Contribution of the header \\

                $P1$ &
                Payload Only &    
                E & 
                E &
                E &
                E &
                E &
                E &
                E &
                E &
                - &
                - &
                Contribution of the payload \\

                \cmidrule(lr){1-13}

                \rowcolor{lightgray}
                $E1$ &
                Encrypted Payload Only &
                E & 
                E &
                E &
                E &
                E &
                E &
                E &
                E &
                ENC  &
                E &
                E. payload's contribution \\

                $E2$ &
                $E1$ - Masked &
                E & 
                E &
                E &
                E &
                E &
                E &
                E &
                E &
                MSK &
                E &
                E. payload's length's contribution \\

                \rowcolor{lightgray}
                $E3$ &
                $E1$ - Obfuscated &
                E & 
                E &
                E &
                E &
                E &
                E &
                E &
                E &
                OBF &
                E &
                Implicit pattern's contribution \\


                
            \bottomrule

            \addlinespace[4pt]
            \multicolumn{13}{l}{ 

                \begin{tabular}[l]{@{}l@{}}

                    \rowcolor{white}
                    
                    \normalsize
                    \textbf{L2}=Ethernet layer; \hspace{0.2cm} 
                    \textbf{L3}=Network layer; \hspace{0.2cm} 
                    \textbf{L4}=Transport layer; \hspace{0.2cm} 
                    \textbf{L7}=Encrypted payload; \hspace{0.2cm} 
                    (Only features in question are shown in the table)
                    \\[2pt]
                    
                    \normalsize
                    \textbf{A.}=Address; \hspace{0.2cm} 
                    \textbf{Ports}=Source and destination ports; \hspace{0.2cm} 
                    \textbf{Seq \& Ack}=Sequence \& Acknowledgment numbers; \hspace{0.2cm} 
                    \textbf{Window S.}=Window size; 
                    \\[2pt]

                    \normalsize 
                    \textbf{O.}=Overfitting; \hspace{0.2cm} 
                    \textbf{E}=Eradicate (replace relevant bytes with $0x00$); \hspace{0.2cm} 
                    \textbf{R}=Randomize (replace relevant bytes with $0xnn$, where $n$ is a random hexadecimal value); \hspace{0.2cm}     
                     \\[2pt]
                    
                    \normalsize 
                    \textbf{ENC}=Encrypted; \hspace{0.2cm} 
                    \textbf{MSK}=Replace encrypted bytes with $0xFF$; \hspace{0.2cm} 
                    \textbf{OBF}=Randomize encrypted bytes with $0xnn$, where $n$ is a random hexadecimal value; \hspace{0.2cm}     
                    \\[2pt]
                    
                \end{tabular}
                
            }
            
        \end{tabular}
    }
    
\end{table*}

We further employ \textit{C} and \textit{T} occlusions to assess contextual and temporal overfitting (see Sections \ref{subsubsec:contextual_overfitting} and \ref{subsubsec:temporal_overfitting}). We randomize context- and time-dependent header features in these occlusions to identify the model’s reliance on these session-specific artifacts. To evaluate the combined impact of all these overfitting shortcuts, the \textit{CTD} occlusion randomizes or removes all features associated with \textit{D1, D2, C} and \textit{T} occlusions.

To examine the distinct roles of header and payload in NTC, we implement \textit{H1} and \textit{P1} occlusions. All header information is preserved in \textit{H1} while the payload is eradicated and padded to a uniform length with $0x00$. This ensures that the model’s reliance on header features alone can be observed without interference from payload variations. Conversely, all header information is removed in \textit{P1}, eliminating all identifying metadata from the headers. We also strip TCP options from the representation to avoid unintended impacts from variable header lengths, allowing the analysis to focus solely on the payload’s contribution to classification performance.

The \textit{E1}, \textit{E2}, and \textit{E3} occlusions assess whether state-of-the-art classifiers can identify patterns in the encrypted payload or if they rely solely on payload length. In \textit{E1}, we isolate the encrypted payload by removing all plaintext information (e.g., header details) to measure its standalone impact. In \textit{E2}, we mask the encrypted payload with $0xFF$, removing inherent randomness and leaving only payload length as a potential feature. Finally, \textit{E3} replaces the encrypted payload with pseudo-random values independent of class labels, simulating \textit{perfect randomness}. This configuration further helps us determine whether any discernible patterns are artifacts of payload length rather than encryption-induced imperfections. All three occlusion techniques were applied to data extraction strategies that do not depend on the \textit{first $m$ bytes} or {first $n$ packets} as they do not exclusively contain encrypted payloads due to handshake packets.

As shown in Table \ref{tab:4_overfitting_vs_granularity}, not all overfitting tendencies affect every traffic granularity. Therefore, we run occlusion experiments only on the relevant design choices. To ensure a fair and unbiased evaluation, we iteratively train and test the selected models for each occlusion separately (as opposed to training once and testing against different occluded data). Drawing on the results presented in Tables \ref{tab:7_occlusion_results_cipherspectrum} and \ref{tab:7_occlusion_results_cstnettls}, Sections \ref{subsubsec:s2rq1} to \ref{subsubsec:s3rq3} offer an in-depth analysis of the experimental findings. However, to maintain conciseness, we report the average accuracy of each occlusion across different design choices. Specifically, for each occlusion type, the accuracy values discussed (e.g., for ET-BERT under A1 occlusion) represent the mean accuracy computed across the 12 design choices presented in Tables \ref{tab:7_occlusion_results_cipherspectrum} and \ref{tab:7_occlusion_results_cstnettls}. 

\subsubsection{S2-RQ1: Data leakage overfitting}
\label{subsubsec:s2rq1}
To address the data leakage concerns discussed in Section \ref{subsec:snag_3_design_choices}, we use the \textit{A1}, \textit{D1}, and \textit{D2} occlusions, as outlined in Table \ref{tab:6_occlusion_strategies}. For \textit{A1}, we conduct 48 experiments across two classifiers and 12 design choices, with results presented in Tables \ref{tab:7_occlusion_results_cipherspectrum} and \ref{tab:7_occlusion_results_cstnettls}. This configuration establishes baseline performance, with ET-BERT and YaTC achieving average accuracies of 0.96 and 0.90, respectively.

\textit{Impact of SII}: To examine the impact of Strong Identification Information (SII), we repeat the 24 experiments for each classifier using the \textit{D1} occlusion. On average, ET-BERT achieved an accuracy of 0.51 ($\downarrow 0.45$), while YaTC reached 0.62 ($\downarrow 0.28$). The average accuracy drop of 0.36 highlights the influence of SII on classifier performance, exposing the risk of overfitting. Besides the resultant poor generalizability, obfuscation practices, such as MAC address randomization and dynamic IP allocation, reduce the reliability of SII, as these features change frequently or can be intentionally altered to protect privacy \cite{sii_random_1, sii_random_2}.

We commend the studies shown in Table \ref{tab:1_lit_summary} that avoid using SII for classification, emphasizing that MAC addresses, IP addresses, and protocol ports should not be relied upon as features for classification.

\textit{Impact of SNI}: The \textit{D2} occlusion is applied to design choices that rely on the initial portion of a TCP/UDP session. Specifically, this includes data extraction strategies that focus on the \textit{First $m$ bytes}, \textit{First $m$ bytes of $n$ packets}, and \textit{First $m$ bytes per packet of $n$ packets}, as shown in Tables \ref{tab:7_occlusion_results_cipherspectrum} and \ref{tab:7_occlusion_results_cstnettls}. When using the \textit{D1} occlusion on these strategies, ET-BERT and YaTC achieved average accuracies of 0.17 and 0.44, respectively. Under \textit{D2}, ET-BERT's accuracy remained largely unchanged at 0.16 ($\downarrow 0.01$), while YaTC’s accuracy decreased to 0.38 ($\downarrow 0.06$). This difference is potentially due to the data representation sizes each classifier uses. ET-BERT represents \textit{T1} and \textit{T2} granularities with 640 bytes and \textit{T3} with 128 bytes per packet across 5 packets, which is unlikely to capture the SNI. In contrast, YaTC uses 1600 bytes for \textit{T1} and \textit{T2} granularities and 320 bytes per packet for 5 packets in \textit{T3}, making it more likely to include the SNI.

The influence of SNI on classification accuracy is minimal in our experiments due to its inconsistent capture across different TLS implementations. The position of the SNI extension within the ClientHello message varies, appearing earlier in Firefox and later in Chromium. Since data extraction strategies operate on fixed-length byte segments, the SNI may or may not be included, showcasing minimal overall impact on accuracy. However, this positional variance highlights the importance of considering implementation-specific factors, as unintended SNI exposure can still introduce bias. As hypothesized in Section \ref{subsubsec:data_leakage_overfitting}, classifier accuracy is affected by SNI data leakage, though the extent of this impact depends on design choices and data representation sizes.

\textit{\textbf{Guideline 1}: Avoid using Strong Identification Information (SII) features such as MAC addresses, IP addresses, and protocol ports, as they contribute to overfitting and reduce model generalizability.}

\textit{\textbf{Guideline 2}: Obfuscate or exclude SNI data in initial portions of sessions to prevent data leakage overfitting, mainly when using large data representations that may capture the SNI unintentionally.}

\begin{table*}[!t]
\renewcommand{\arraystretch}{0.7}

    \caption{Feature occlusion results - Classification accuracies against CipherSpectrum dataset}
    \label{tab:7_occlusion_results_cipherspectrum}
    
    \centering
    \small

    \resizebox{0.95\textwidth}{!}{        
        \begin{tabular}{cc*{12}{C{1cm}}}

            \toprule

                \multirow{2}{*}{\textbf{ID}} &
                \multirow{2}{*}{\textbf{Model}} &
                \multirow{2}{*}{\textbf{Packet}} &
                \multirow{2}{*}{\textbf{Burst}} &
                \multicolumn{5}{c}{\textbf{Flow}} &
                \multicolumn{5}{c}{\textbf{Session}}  \\

                \cmidrule(lr){5-9} 
                \cmidrule(lr){10-14}

                \rowcolor{white}
                &
                &
                &
                &
                
                \textbf{T1}&
                \textbf{T2 \ding{72}}&
                \textbf{T2 \ding{68}}&
                \textbf{T3 \ding{72}}&
                \textbf{T3 \ding{68}}&

                \textbf{T1}&
                \textbf{T2 \ding{72}}&
                \textbf{T2 \ding{68}}&
                \textbf{T3 \ding{72}}&
                \textbf{T3 \ding{68}}
                \\
            
            \midrule

                    \rowcolor{lightgray}
                    \multirow{2}{*}{A1} &
                    ET-BERT &
                    0.99 &
                    0.98 &
                    0.96 &
                    0.96 &
                    0.97 &
                    0.90 &
                    0.98 &
                    0.93 &
                    0.94 &
                    0.96 &
                    0.95 &
                    0.99
                    \\
    
                    \cmidrule(lr){2-14}
    
                    &
                    YaTC &
                    0.94 &
                    0.90 &
                    0.84 &
                    0.84 &
                    0.80 &
                    0.87 &
                    0.89 &
                    0.86 &
                    0.89 &
                    0.80 &
                    0.90 &
                    0.91
                    \\

                \cmidrule(lr){1-14}
                
                    \rowcolor{lightgray}
                    \multirow{2}{*}{D1} &
                    ET-BERT &
                    0.79 &
                    0.41 &
                    0.10 &
                    0.12 &
                    0.31 &
                    0.08 &
                    0.29 &
                    0.11 &
                    0.10 &
                    0.23 &
                    0.57 &
                    0.79
                    \\
    
                    \cmidrule(lr){2-14}
    
                    &
                    YaTC &
                    0.42 &
                    0.74 &
                    0.51 &
                    0.49 &
                    0.36 &
                    0.54 &
                    0.40 &
                    0.36 &
                    0.36 &
                    0.31 &
                    0.42 &
                    0.41
                    \\

                \cmidrule(lr){1-14}
                
                    \rowcolor{lightgray}
                    \multirow{2}{*}{D2} &
                    ET-BERT &
                    - &
                    - &
                    0.08 &
                    0.10 &
                    - &
                    0.08 &
                    - &
                    0.10 &
                    0.09 &
                    - &
                    0.52 &
                    -
                    \\
    
                    \cmidrule(lr){2-14}
    
                    &
                    YaTC &
                    - &
                    - &
                    0.43 &
                    0.38 &
                    - &
                    0.50 &
                    - &
                    0.34 &
                    0.29 &
                    - &
                    0.37 &
                    -
                    \\

                \cmidrule(lr){1-14}
                
                    \rowcolor{lightgray}
                    \multirow{2}{*}{C} &
                    ET-BERT &
                    0.68 &
                    0.35 &
                    - &
                    - &
                    0.29 &
                    - &
                    0.20 &
                    - &
                    - &
                    0.18 &
                    - &
                    0.68
                    \\
    
                    \cmidrule(lr){2-14}
    
                    &
                    YaTC &
                    0.37 &
                    0.69 &
                    - &
                    - &
                    0.35 &
                    - &
                    0.33 &
                    - &
                    - &
                    0.30 &
                    - &
                    0.37
                    \\

                \cmidrule(lr){1-14}
                
                    \rowcolor{lightgray}
                    \multirow{2}{*}{T} &
                    ET-BERT &
                    0.76 & 
                    0.38 &
                    - &
                    - &
                    0.30 &
                    - &
                    0.23 &
                    - &
                    - &
                    0.21 &
                    - &
                    0.71
                    \\
    
                    \cmidrule(lr){2-14}
    
                    &
                    YaTC &
                    0.40 &
                    0.73 &
                    - &
                    - &
                    0.36 &
                    - &
                    0.38 &
                    - &
                    - &
                    0.29 &
                    - &
                    0.39
                    \\

                \cmidrule(lr){1-14}
                
                    \rowcolor{lightgray}
                    \multirow{2}{*}{CTD} &
                    ET-BERT &
                    0.62 & 
                    0.33 &
                    - &
                    - &
                    0.26 &
                    - &
                    0.19 &
                    - &
                    - &
                    0.17 &
                    - &
                    0.63
                    \\
    
                    \cmidrule(lr){2-14}
    
                    &
                    YaTC &
                    0.33 &
                    0.69 &
                    - &
                    - &
                    0.35 &
                    - &
                    0.32 &
                    - &
                    - &
                    0.28 &
                    - &
                    0.33
                    \\
                    
                \cmidrule(lr){1-14}
                
                    \rowcolor{lightgray}
                    \multirow{2}{*}{H1} &
                    ET-BERT &
                    0.80 &
                    0.37 &
                    0.10 &
                    0.09 &
                    0.29 &
                    0.10 &
                    0.28 &
                    0.10 &
                    0.12 &
                    0.25 &
                    0.58 &
                    0.79
                    \\
    
                    \cmidrule(lr){2-14}
    
                    &
                    YaTC &
                    0.44 &
                    0.72 &
                    0.27 &
                    0.27 &
                    0.22 &
                    0.48 &
                    0.28 &
                    0.11 &
                    0.26 &
                    0.23 &
                    0.30 &
                    0.36
                    \\

                
    
    
                
                \cmidrule(lr){1-14}
                
                    \rowcolor{lightgray}
                    \multirow{2}{*}{E1} &
                    ET-BERT &
                    0.12 &
                    0.12 &
                    - &
                    - &
                    0.14 &
                    - &
                    0.13 &
                    - &
                    - &
                    0.11 &
                    - &
                    0.11
                    \\
    
                    \cmidrule(lr){2-14}
    
                    &
                    YaTC &
                    0.30 &
                    0.31 &
                    - &
                    - &
                    0.29 &
                    - &
                    0.26 &
                    - &
                    - &
                    0.28 &
                    - &
                    0.25
                    \\

                \cmidrule(lr){1-14}
                
                    \rowcolor{lightgray}
                    \multirow{2}{*}{E2} &
                    ET-BERT &
                    0.12 &
                    0.13 &
                    - &
                    - &
                    0.14 &
                    - &
                    0.13 &
                    - &
                    - &
                    0.11 &
                    - &
                    0.11
                    \\
    
                    \cmidrule(lr){2-14}
    
                    &
                    YaTC &
                    0.36 &
                    0.39 &
                    - &
                    - &
                    0.42 &
                    - &
                    0.37 &
                    - &
                    - &
                    0.39 &
                    - &
                    0.33
                    \\

                \cmidrule(lr){1-14}
                
                    \rowcolor{lightgray}
                    \multirow{2}{*}{E3} &
                    ET-BERT &
                    0.11 &
                    0.12 &
                    - &
                    - &
                    0.14 &
                    - &
                    0.13 &
                    - &
                    - &
                    0.10 &
                    - &
                    0.11
                    \\
    
                    \cmidrule(lr){2-14}
    
                    &
                    YaTC &
                    0.31 &
                    0.30 &
                    - &
                    - &
                    0.31 &
                    - &
                    0.27 &
                    - &
                    - &
                    0.28 &
                    - &
                    0.26
                    \\

                \cmidrule(lr){1-14}
                
                    \rowcolor{lightgray}
                    \multirow{2}{*}{$\hat{E2}$} &
                    ET-BERT &
                    0.12 &
                    0.12 &
                    - &
                    - &
                    0.11 &
                    - &
                    0.12 &
                    - &
                    - &
                    0.13 &
                    - &
                    0.12 
                    \\
    
                    \cmidrule(lr){2-14}
    
                    &
                    YaTC &
                    0.34 &
                    0.32 &
                    - &
                    - &
                    0.37 &
                    - &
                    0.33 &
                    - &
                    - &
                    0.35 &
                    - &
                    0.3
                    \\

                \cmidrule(lr){1-14}
                
                    \rowcolor{lightgray}
                    \multirow{2}{*}{$\hat{\hat{E2}}$} &
                    ET-BERT &
                    0.12 &
                    0.11 &
                    - &
                    - &
                    0.12 &
                    - &
                    0.11 &
                    - &
                    - &
                    0.13 &
                    - &
                    0.11  
                    \\
    
                    \cmidrule(lr){2-14}
    
                    &
                    YaTC &
                    0.31 &
                    0.29 &
                    - &
                    - &
                    0.34 &
                    - &
                    0.3 &
                    - &
                    - &
                    0.29 &
                    - &
                    0.27
                    \\
                    
            \bottomrule




                    
                    
                
            
        \end{tabular}
    }
    
\end{table*}
\begin{table*}[!t]
\renewcommand{\arraystretch}{0.7}

    \caption{Feature occlusion results - Classification accuracies against CSTNET-TLS1.3 dataset}
    \label{tab:7_occlusion_results_cstnettls}
    
    \centering
    \small

    \resizebox{0.95\textwidth}{!}{        
        \begin{tabular}{cc*{12}{C{1cm}}}

            \toprule

                \multirow{2}{*}{\textbf{ID}} &
                \multirow{2}{*}{\textbf{Model}} &
                \multirow{2}{*}{\textbf{Packet}} &
                \multirow{2}{*}{\textbf{Burst}} &
                \multicolumn{5}{c}{\textbf{Flow}} &
                \multicolumn{5}{c}{\textbf{Session}}  \\

                \cmidrule(lr){5-9} 
                \cmidrule(lr){10-14}

                \rowcolor{white}
                &
                &
                &
                &
                
                \textbf{T1}&
                \textbf{T2 \ding{72}}&
                \textbf{T2 \ding{68}}&
                \textbf{T3 \ding{72}}&
                \textbf{T3 \ding{68}}&

                \textbf{T1}&
                \textbf{T2 \ding{72}}&
                \textbf{T2 \ding{68}}&
                \textbf{T3 \ding{72}}&
                \textbf{T3 \ding{68}}
                \\
            
            \midrule

                    \rowcolor{lightgray}
                    \multirow{2}{*}{A1} &
                    ET-BERT &
                    0.99 &
                    0.98 &
                    0.95 &
                    0.96 &
                    0.99 &
                    0.95 &
                    0.99 &
                    0.93 &
                    0.92 &
                    0.99 &
                    0.92 &
                    0.98
                    \\
    
                    \cmidrule(lr){2-14}
    
                    &
                    YaTC &
                    0.98 &
                    0.83 &
                    0.89 &
                    0.90 &
                    0.93 &
                    0.94 &
                    0.96 &
                    0.95 &
                    0.95 &
                    0.93 &
                    0.96 &
                    0.97
                    \\

                \cmidrule(lr){1-14}
                
                    \rowcolor{lightgray}
                    \multirow{2}{*}{D1} &
                    ET-BERT &
                    0.97 &
                    0.74 &
                    0.50 &
                    0.53 &
                    0.79 &
                    0.57 &
                    0.79 &
                    0.66 &
                    0.69 &
                    0.76 &
                    0.71 &
                    0.73
                    \\
    
                    \cmidrule(lr){2-14}
    
                    &
                    YaTC &
                    0.90 &
                    0.61 &
                    0.81 &
                    0.79 &
                    0.70 &
                    0.80 &
                    0.81 &
                    0.92 &
                    0.92 &
                    0.68 &
                    0.95 &
                    0.86
                    \\

                \cmidrule(lr){1-14}
                
    
    

                
                    \rowcolor{lightgray}
                    \multirow{2}{*}{C} &
                    ET-BERT &
                    0.96 &
                    0.75 &
                    - &
                    - &
                    0.72 &
                    - &
                    0.68 &
                    - &
                    - &
                    0.74 &
                    - &
                    0.71
                    \\
    
                    \cmidrule(lr){2-14}
    
                    &
                    YaTC &
                    0.86 &
                    0.58 &
                    - &
                    - &
                    0.66 &
                    - &
                    0.75 &
                    - &
                    - &
                    0.63 &
                    - &
                    0.81
                    \\

                \cmidrule(lr){1-14}
                
                    \rowcolor{lightgray}
                    \multirow{2}{*}{T} &
                    ET-BERT &
                    0.94 &
                    0.68 &
                    - &
                    - &
                    0.64 &
                    - &
                    0.63 &
                    - &
                    - &
                    0.70 &
                    - &
                    0.68
                    \\
    
                    \cmidrule(lr){2-14}
    
                    &
                    YaTC &
                    0.86 &
                    0.58 &
                    - &
                    - &
                    0.61 &
                    - &
                    0.75 &
                    - &
                    - &
                    0.62 &
                    - &
                    0.83
                    \\

                \cmidrule(lr){1-14}
                
                    \rowcolor{lightgray}
                    \multirow{2}{*}{CTD} &
                    ET-BERT &
                    0.83 & 
                    0.67 &
                    - &
                    - &
                    0.49 &
                    - &
                    0.51 &
                    - &
                    - &
                    0.61 &
                    - &
                    0.56
                    \\
    
                    \cmidrule(lr){2-14}
    
                    &
                    YaTC &
                    0.69 &
                    0.52 &
                    - &
                    - &
                    0.51 &
                    - &
                    0.63 &
                    - &
                    - &
                    0.45 &
                    - &
                    0.73
                    \\
                    
                \cmidrule(lr){1-14}
                
                    \rowcolor{lightgray}
                    \multirow{2}{*}{H1} &
                    ET-BERT &
                    0.96 &
                    0.77 &
                    0.68 &
                    0.69 &
                    0.78 &
                    0.58 &
                    0.79 &
                    0.81 &
                    0.79 &
                    0.78 &
                    0.72 &
                    0.77
                    \\
    
                    \cmidrule(lr){2-14}
    
                    &
                    YaTC &
                    0.90 &
                    0.62 &
                    0.68 &
                    0.68 &
                    0.73 &
                    0.69 &
                    0.80 &
                    0.80 &
                    0.79 &
                    0.71 &
                    0.86 &
                    0.85
                    \\

                
    
    
                
                \cmidrule(lr){1-14}
                
                    \rowcolor{lightgray}
                    \multirow{2}{*}{E1} &
                    ET-BERT &
                    0.13 &
                    0.13 &
                    - &
                    - &
                    0.11 &
                    - &
                    0.11 &
                    - &
                    - &
                    0.11 &
                    - &
                    0.09
                    \\
    
                    \cmidrule(lr){2-14}
    
                    &
                    YaTC &
                    0.38 &
                    0.34 &
                    - &
                    - &
                    0.35 &
                    - &
                    0.26 &
                    - &
                    - &
                    0.37 &
                    - &
                    0.22
                    \\

                \cmidrule(lr){1-14}
                
                    \rowcolor{lightgray}
                    \multirow{2}{*}{E2} &
                    ET-BERT &
                    0.13 &
                    0.14 &
                    - &
                    - &
                    0.12 &
                    - &
                    0.11 &
                    - &
                    - &
                    0.11 &
                    - &
                    0.10
                    \\
    
                    \cmidrule(lr){2-14}
    
                    &
                    YaTC &
                    0.46 &
                    0.40 &
                    - &
                    - &
                    0.47 &
                    - &
                    0.33 &
                    - &
                    - &
                    0.50 &
                    - &
                    0.34
                    \\

                \cmidrule(lr){1-14}
                
                    \rowcolor{lightgray}
                    \multirow{2}{*}{E3} &
                    ET-BERT &
                    0.13 &
                    0.14 &
                    - &
                    - &
                    0.12 &
                    - &
                    0.11 &
                    - &
                    - &
                    0.11 &
                    - &
                    0.10
                    \\
    
                    \cmidrule(lr){2-14}
    
                    &
                    YaTC &
                    0.38 &
                    0.33 &
                    - &
                    - &
                    0.35 &
                    - &
                    0.27 &
                    - &
                    - &
                    0.37 &
                    - &
                    0.22
                    \\

                \cmidrule(lr){1-14}
                
                    \rowcolor{lightgray}
                    \multirow{2}{*}{$\hat{E2}$} &
                    ET-BERT &
                    0.11 &
                    0.13 &
                    - &
                    - &
                    0.12 &
                    - &
                    0.12 &
                    - &
                    - &
                    0.13 &
                    - &
                    0.11 
                    \\
    
                    \cmidrule(lr){2-14}
    
                    &
                    YaTC &
                    0.24 &
                    0.39 &
                    - &
                    - &
                    0.47 &
                    - &
                    0.31 &
                    - &
                    - &
                    0.48 &
                    - &
                    0.33
                    \\

                \cmidrule(lr){1-14}
                
                    \rowcolor{lightgray}
                    \multirow{2}{*}{$\hat{\hat{E2}}$} &
                    ET-BERT &
                    0.12 &
                    0.14 &
                    - &
                    - &
                    0.13 &
                    - &
                    0.11 &
                    - &
                    - &
                    0.12 &
                    - &
                    0.13 
                    \\
    
                    \cmidrule(lr){2-14}
    
                    &
                    YaTC &
                    0.22 &
                    0.35 &
                    - &
                    - &
                    0.37 &
                    - &
                    0.28 &
                    - &
                    - &
                    0.39 &
                    - &
                    0.31
                    \\
                
            \bottomrule

            \addlinespace[8pt]
            \multicolumn{13}{l}{ 

                \begin{tabular}[l]{@{}l@{}}

                    \rowcolor{white}

                    \ding{72}=First $n$ Packets; \hspace{0.2cm} 
                    \ding{68}=Any consecutive $n$ Packets;\\[3pt]
                    
                    \textbf{T1}=Type 1(First $m$ bytes); \hspace{0.2cm} 
                    \textbf{T2}=Type 2(First $m$ bytes of $n$ packets); \hspace{0.2cm}
                    \textbf{T3}=Type 3(First $m$ bytes per packet of $n$ Packets); \hspace{0.2cm} \\[3pt]
                    
                \end{tabular}
                
            }
            
        \end{tabular}
    }
    
\end{table*}

\subsubsection{S2-RQ2: Contextual overfitting}
To assess the presence of contextual overfitting, we implemented the \textit{C} occlusion, which randomizes session-specific fields such as the IP Identification (IP ID), IP header checksum, and TCP sequence and acknowledgment numbers. 

As discussed in Section \ref{subsubsec:contextual_overfitting}, contextual overfitting occurs when a single TCP/UDP session is split into multiple samples used in training and testing sets. To demonstrate, we focused on packet and burst-based granularities and employed data extraction strategies that select \textit{any consecutive $n$ packets}, as these are more susceptible to contextual overfitting.

Our experimental results, summarized in Tables \ref{tab:7_occlusion_results_cipherspectrum} and \ref{tab:7_occlusion_results_cstnettls}, show that under the \textit{C} occlusion, the average accuracies of ET-BERT and YaTC decreased to 0.57 ($\downarrow 0.06$) and 0.55 ($\downarrow 0.05$), respectively. This is a noticeable drop from their average accuracies of 0.63 and 0.60 under the baseline condition \textit{D1}.

The consistent decline in accuracy across all evaluated design choices when applying the \textit{C} occlusion indicates that the models rely on session-specific contextual features as shortcuts for classification.

This reduction in performance highlights the importance of mitigating context-dependent, session-specific artifacts in the dataset.

\textit{\textbf{Guideline 3}: Avoid uninformative, session-specific fields such as IP ID, IP header checksum, and TCP sequence/acknowledgment numbers to reduce contextual overfitting risks.}

\subsubsection{S2-RQ3: Temporal overfitting}
We applied the \textit{T} occlusion strategy to assess temporal overfitting, which involves randomizing time-variant fields such as the TCP option timestamps.

Mirroring our approach for contextual overfitting, we conducted 12 experiments for each model, as reflected in Tables \ref{tab:7_occlusion_results_cipherspectrum} and \ref{tab:7_occlusion_results_cstnettls}. Consistent with the previous findings, both ET-BERT and YaTC showed a decrease in average accuracy under the \textit{T} occlusion. Their accuracies dropped to 0.57 ($\downarrow 0.06$) and 0.56 ($\downarrow 0.04$), respectively, compared to their baseline accuracies of 0.63 and 0.60.

This decline in performance indicates that the models were leveraging time-dependent features as shortcuts for classification, confirming the presence of temporal overfitting. 

\textit{\textbf{Guideline 4}: Exclude or randomize time-variant fields, such as TCP timestamps, to prevent models from developing dependencies on temporal patterns.}

\subsubsection{Consequences of Design Choice Oversights}
To further illustrate the impact of design choice oversights, we evaluated ET-BERT and YaTC using their originally proposed data preprocessing and representation techniques (as opposed to preprocessing methods discussed in Appendix \ref{app_sec:data_preprocessing}). Under these original conditions, ET-BERT and YaTC achieved high average accuracies of 0.91 and 0.93 across the two datasets. However, when the features associated with data leakage, contextual overfitting, and temporal overfitting were randomized, the accuracy of ET-BERT dropped sharply to 0.59 ($\downarrow 0.32$), while YaTC’s accuracy fell to 0.68 ($\downarrow 0.25$).

These significant accuracy declines underscore the extent to which both models relied on features prone to data leakage and overfitting. This highlights the need for careful design choices to avoid reliance on artifacts that compromise the model's generalizability and robustness.

\textit{\textbf{Guideline 5}: Ensure data extraction strategies minimize reliance on contextual and temporal features by avoiding overlap between training and testing samples drawn from the same sessions.}

\subsubsection{S3-RQ1: Patterns in Cipher Texts}
In this section, we investigate the existence of discernible patterns in encrypted payloads, some of which claim to be due to imperfect randomness in encryption ciphers. ET-BERT and YaTC have explicitly suggested their ability to exploit patterns in encrypted payloads, serving as ideal candidates for this analysis.

First, we perform the \textit{E1} analysis, as outlined in Table \ref{tab:6_occlusion_strategies}. In this configuration, we use only the encrypted payload, removing any plaintext information (e.g., header details) to isolate its contribution to classification. As shown in Tables \ref{tab:7_occlusion_results_cipherspectrum} and \ref{tab:7_occlusion_results_cstnettls}, we conducted 12 experiments per classifier and on average, ET-BERT achieved an accuracy of 0.12, while YaTC achieved an accuracy of 0.30.

Next, we apply a masking strategy in the \textit{E2} occlusion configuration, replacing the encrypted payload bytes with $0xFF$. This test addresses the hypothesis of \textit{imperfect randomness} by removing any underlying randomness in the payload, leaving only the encrypted payload length as a potential feature for the model. Under these conditions, ET-BERT maintained an accuracy of 0.12, while YaTC’s accuracy increased to 0.39 ($\uparrow 0.09$).

The increase in classifier accuracy under the \textit{E2} configuration can be attributed to differences in data representation and the impact of masking, which removes noise introduced by encryption. ET-BERT processes network traffic as a token sequence without explicit packet boundaries, limiting its ability to leverage structural information. In contrast, YaTC’s Matrix Flow Representation organizes traffic into structured segments, preserving per-packet boundaries. This structured approach allows YaTC to better exploit class-specific patterns in payload lengths that remain observable in TLS 1.3 despite encryption \cite{tls_13}. By eliminating encryption-induced randomness, the \textit{E2} masking strategy enhances YaTC’s ability to recognize these patterns, leading to improved classification accuracy.

Finally, in the \textit{E3} occlusion configuration, we replace the encrypted payload bytes with pseudo-random values independent of the class labels, simulating a scenario of \textit{perfect randomness}. This approach aims to eliminate any patterns that could arise from encryption, further testing whether observable patterns stem from payload length rather than cipher-related randomness. In this setup, ET-BERT again averaged an accuracy of 0.12, and YaTC maintained an accuracy of 0.30.

Collectively, the 72 experiments presented in Tables \ref{tab:7_occlusion_results_cipherspectrum} and \ref{tab:7_occlusion_results_cstnettls} demonstrate that state-of-the-art classifiers do not learn any intrinsic patterns from encrypted payloads beyond their length. This finding aligns with the guarantees provided by TLS 1.3, which asserts that the only observable characteristic in encrypted payloads is their length \cite{tls_13}. Thus, our results reinforce that any previously perceived patterns within encrypted payloads are likely artifacts from outdated datasets containing unencrypted data.

\textit{\textbf{Guideline 6}: Focus on encrypted payload length rather than content, as classifiers primarily rely on payload length for classification rather than intrinsic patterns in the cipher text.}

\textit{\textbf{Guideline 7}: Use structured representations (e.g., segmented matrices) to better capture payload length variations, especially when packet boundaries are relevant to the model.}

\textit{\textbf{Guideline 8}: Be cautious in interpreting patterns observed in encrypted data, as they may result from unencrypted or outdated datasets rather than meaningful insights in modern encrypted traffic.}

\subsubsection{S3-RQ2: Sufficiency of Encrypted Payload Alone}
To evaluate whether encrypted payload alone is sufficient for classification, we applied the \textit{D1}, \textit{H1}, and \textit{E1} occlusions, as outlined in Table \ref{tab:6_occlusion_strategies}. For reference, ET-BERT and YaTC achieved baseline accuracies of 0.63 and 0.60 under the \textit{D1} condition. When isolated to encrypted payload only (\textit{E1} occlusion), ET-BERT’s accuracy dropped significantly to 0.12 ($\downarrow 0.51$), and YaTC’s to 0.30 ($\downarrow 0.30$), suggesting that encrypted payload alone provides limited information for accurate classification.

Next, we isolated header information using the \textit{H1} occlusion. ET-BERT matched its baseline \textit{D1} accuracy of 0.63, indicating that its classification is entirely based on header information. In contrast, YaTC’s accuracy under \textit{H1} dropped slightly to 0.57 ($\downarrow 0.03$), showing that it relies on both header and payload information for classification, though header information plays a dominant role.

These results suggest that claims regarding the sufficiency of encrypted payload alone for classification may stem from artifacts present in unencrypted network traffic datasets. While encrypted payload lengths contribute when appropriately represented, they are insufficient for accurate classification in contemporary networks where headers remain essential.

\textit{\textbf{Guideline 9}: Avoid relying solely on encrypted payloads for classification, as they provide limited information; header data remains crucial for accuracy.}

\textit{\textbf{Guideline 10}: Be cautious of assumptions about payload sufficiency, as these may originate from artifacts in outdated or unencrypted datasets.}

\textit{\textbf{Guideline 11}: Leverage both header and payload length information to improve classification performance, especially in modern, encrypted network environments.}

\subsubsection{S3-RQ3: Truncating, padding vs performance}
\label{subsubsec:s3rq3}
To further examine the role of payload length in classification, we tested the impact of truncating and padding the payload. Recognizing that manipulating the only learnable characteristic—payload length—can reduce performance, we implemented two occlusion strategies: $\hat{E2}$ and $\hat{\hat{E2}}$.

In $\hat{E2}$, we truncate the payload by 25\%, representing the traffic with only $0.75m$ bytes (where $m$ is the original byte length used for the granularity). Similarly, in $\hat{\hat{E2}}$, we truncate the payload by 50\%, using $0.5m$ bytes to represent traffic. For YaTC, additional padding of 25\% and 50\% is applied, respectively, to compensate for the reduced payload length. However, this padding adjustment does not impact ET-BERT due to its token sequence-based representation.

Compared to the baseline \textit{E2}, YaTC’s average accuracy decreased from 0.39 to 0.35 ($\downarrow 0.04$) and 0.31 ($\downarrow 0.08$) for ${\hat{E2}}$ and $\hat{\hat{E2}}$, highlighting its dependency on payload length for classification. In contrast, ET-BERT’s average accuracy remained unchanged, further demonstrating that its linguistic representation does not leverage the encrypted payloads' length for classification. 

As RFC 8446 states, "\textit{TLS does not hide the length of the data it transmits, though endpoints can pad TLS records to obscure lengths.}" \cite{tls_13}. These experiments highlight the effects of truncation and padding on classification performance, reinforcing the significance of payload length as a critical feature in NTC. 

\textit{\textbf{Guideline 12}: Avoid arbitrary truncation or padding of payloads, as these modifications can significantly impact classifier accuracy. Especially for models that rely on payload length as a key feature.}

\section{Discussion}
This study systematically examined the current landscape of raw-information-based network traffic classification (NTC) identifying key challenges that impact its effectiveness. Our findings highlight the importance of prioritizing updated datasets that reflect modern encryption protocols. Similarly, we emphasize that design choices and features should be selected based on a thorough evaluation of RFCs (Request for Comments) to assess suitability and informativeness. Future work could broaden the scope of analysis to explore a broader range of models and datasets. Furthermore, we highlight the need for greater transparency and reproducibility in developing ML models for NTC. 

\textbf{Limited Scope}: Researchers, particularly within the network security community, have emphasized the critical need for improved reproducibility \cite{reproducability, dl_problems_2_trustee}, a call to action we support. As discussed in Section \ref{subsubsec:model_selection}, the scope of our analysis was confined to two fully reproducible models considered state-of-the-art and publicly available datasets. We emphasize that the focus of our analysis was not to evaluate the performance of classifiers but rather to uncover potential pitfalls in design choices. To achieve this, we focused on state-of-the-art models and ensured they represented a diverse range of design choices, enabling a representation of the challenges in the field. Although this may not capture the entirety of practices in the field, our reported findings expose pitfalls that are symptomatic of the broader literature. To maintain transparency and foster further research, we make all scripts related to our evaluation publicly available\footnote{https://github.com/nime-sha256/ntc-enigma}. In this context, our work contributes to ongoing efforts that advocate for a more critical evaluation of developments in this area \cite{dl_problems_2_trustee, dl_problems_1, dl_problems_3}.

\textbf{Validity of CipherSpectrum}: Synthetic data collected using automated scripts without human involvement has faced scrutiny regarding its credibility and representativeness. However, CipherSpectrum addresses the immediate need for a contemporary dataset while remaining valid for evaluating raw-information-based NTC models designed for TLS 1.3 encrypted traffic. This validity stems from the dataset's inclusion of accurate header and payload information, verified by ensuring full rendering of web pages during data collection. In future work, we plan to extend CipherSpectrum by incorporating data collected with human involvement, enhancing its fidelity further. Additionally, we make our instrumented Chromium version publicly available to support and facilitate research in this domain\footnote{https://github.com/nime-sha256/chromium-cipher-suite-customizer}.

\section{Conclusion}
Our systematization of knowledge uncovered drawbacks in raw information-based network traffic classification (NTC). We identified a widespread reliance on outdated datasets, oversights in design choices leading to overfitting, and the consequences of unsubstantiated assumptions.

Specifically, we demonstrated that popularly used datasets include substantial amounts of unencrypted network traffic and do not reflect contemporary security protocols and standards (e.g., TLS 1.3). To address this issue, we introduced CipherSpectrum, a contemporary dataset that embodies modern encryption practices, enabling the development of robust and generalizable NTC models.

Through 348 feature occlusion experiments on state-of-the-art classifiers, we achieved two main objectives: (1) we demonstrated how design oversights can lead to overfitting, hindering the classifiers' generalizability; (2) we validated/refuted prevailing assumptions by providing empirical evidence, thereby reducing confusion and inconsistency in the field.

Building upon our findings, we provided strategic insights and best practices to mitigate the identified issues.

In conclusion, our work underscores the necessity for updated datasets and careful design choices to make NTC more robust and applicable. Further, by reevaluating prevailing assumptions, we paved the way for more effective network traffic classification in today's encrypted landscape.



\ifCLASSOPTIONcompsoc
  \section*{Acknowledgments}
\else
  \section*{Acknowledgment}
\fi

We sincerely thank the reviewers and our shepherd for their constructive feedback, which improved the quality of this work. We also extend our gratitude to Distinguished Professor Yvo Desmedt, Associate Professor Gustavo Batista, and Dr Mohammad Goudarzi for their valuable insights throughout this research.

\newpage
\begingroup
\setstretch{0.9}

\bibliographystyle{IEEEtranS}
\bibliography{IEEEabrv, references}

\begin{thebibliography}{10}
\bibitem{alexa}
E.~Kovacs, ``{A}mazon’s {S}huttering of {A}lexa {R}anking {S}ervice {H}its {C}ybersecurity {I}ndustry --- securityweek.com.''
\end{thebibliography}

\endgroup

\newpage
\appendices

\section{Cipherspectrum}
\label{app_sec:cipherspectrum}

\subsection{Collection Methodology}
Our primary objective is to capture encrypted network traffic and facilitate the downstream task of \textit{web traffic classification} (see Section \ref{subsec:preliminries_tasks_and_datasets}). Therefore, to create CipherSpectrum, we focused on collecting web traffic encrypted with the three cipher suites mandated/strongly-recommended by TLS 1.3: TLS-AES-256-GCM-SHA384, TLS-CHACHA20-POLY1305-SHA256, TLS-AES-128-GCM-SHA256 \cite{tls_13}. This strategy ensures that the dataset aligns with modern encryption standards and contemporary network traffic. 

\subsubsection{Browser Selection and Customization} To browse websites and generate network traffic, we chose two open-source web browsers: Firefox \cite{firefox} and Chromium \cite{chromium}. Firefox inherently allows users to configure cipher suite preferences, enabling us to modify the encryption algorithms used in secure communications directly. However, since Chromium does not provide this capability natively, we customized its source code to add a feature for selecting specific cipher suites. This browser selection and customization enabled us to create a controlled and diverse environment for generating traffic.

\subsubsection{Domain Selection and Automation} To generate realistic web traffic, we selected the top 2000 domains listed on Cloudflare Radar \cite{radar} to align with recent trends and reliable traffic sources following the practice of \cite{foundation_et_bert, data_collection_ccs18, data_collection_ccs14}\footnote{We opted for Cloudflare Radar instead of currently dormant Alexa \cite{alexa} rankings.}. After identifying the top domains, we implemented an automated process to verify their accessibility as follows: if a domain is accessible via a web browser without errors, we extract five URLs belonging to web-pages of the same domain. For example, for the domain "example.com", the extracted URLs would include: https://example.com, https://example.com/contact-us, https://example.com/settings, https://example.com/privacy, and https://example.com/about-us. Any domain that was either inaccessible via a browser or lacked five URLs from the same domain was excluded from the final list. We selected five web pages per domain to capture a relatively diverse set of traffic patterns, and to include a variety of content types within each domain. This filtering process resulted in a refined set of 132 domains and a total of 660 URLs.

\subsubsection{Traffic Collection Process} Further adhering methodologies from previous studies \cite{data_collection_ccs18, data_collection_pets16, data_collection_ccs14, selenium_based_data_collection_1, selenium_based_data_collection_2, selenium_based_data_collection_3, selenium_based_data_collection_4, selenium_based_data_collection_5}, we automated the traffic collection process using Selenium \cite{selenium_web_driver} as shown in Algorithm \ref{alg:2_data_collection}. 

\begin{algorithm}[tb]
    \caption{Network traffic collection} 
    \label{alg:2_data_collection}

    {
        \setstretch{1.1}
        
        \begin{algorithmic}[1]
            \State \textit{urls} $\leftarrow$ load 660 urls of shortlisted domains
            
            \For{each \textit{iteration} in \{1 to 100\}}
                \For{each \textit{c\_suite} in \{"c20", "a128", "a256"\}}
                    \For{each \textit{browser} in \{"chromium", "firefox"\}}
                        \For{each \textit{url} in \textit{urls}}
                            \State run \textit{browser} with \textit{c\_suite}

                            \State start traffic capture with tshark

                            \State load \textit{url} with \textit{browser}

                            \State capture screenshot after page load

                            \State end traffic capture with tshark
    
                            \State quit browser    
                        \EndFor
                    \EndFor
                \EndFor
            \EndFor
        \end{algorithmic}
    }
\end{algorithm}

For each iteration of the traffic collection process, the algorithm begins by iterating over the three target cipher suites: TLS-CHACHA20-POLY1305-SHA256 ("c20"), TLS-AES-128-GCM-SHA256 ("a128"), and TLS-AES-256-GCM-SHA384 ("a256") \textit{(Line 3)}. This ensures that traffic is collected in a balanced manner for all three cipher suites mandated/strongly-recommended by TLS 1.3. Within each iteration, the algorithm proceeds to loop through two selected browsers, Chromium and Firefox, to generate diverse traffic patterns \textit{(Line 4)}.

For each URL in the set of 660 shortlisted URLs \textit{(Line 5)}, the algorithm follows a systematic process. First, the browser is launched with the specified cipher suite for the current iteration \textit{(Line 6)}, ensuring that all traffic generated in this session uses the intended encryption algorithm. Traffic capture then begins using Tshark \textit{(Line 7)}, which records all incoming and outgoing network packets. The algorithm then directs the browser to load the selected URL \textit{(Line 8)}, simulating a user accessing the webpage and generating realistic network traffic. Once the webpage is fully loaded, a screenshot is taken to document the state of the page during the session \textit{(Line 9)}.

After this, the algorithm ends the traffic capture process and saves the recorded packets for analysis \textit{(Line 10)}. Finally, the browser is quit to clean up session data and free up system resources \textit{(Line 11)}. This entire sequence is repeated for each URL \textit{(Line 5)} across each browser \textit{(Line 4)} and each cipher suite \textit{(Line 3)} within every iteration \textit{(Line 2)}. By repeating this procedure across 100 iterations, the algorithm effectively captures a diverse dataset, reflecting traffic encrypted with different cipher suites in varied browser environments.

The sequential website visits allowed for capturing different variants of each site, resulting in a more comprehensive and representative traffic patterns \cite{data_collection_ccs14, data_collection_ccs18}. All data was collected using a university network, providing a realistic environment from January 13, 2024, to March 10, 2024.

\subsection{Traffic Lebelling}
At the conclusion of the traffic collection process, we successfully gathered a total of 396,000 traffic traces, resulting from 100 iterations across three different cipher suites, two browsers, and 660 URLs. To prepare this raw traffic data for analysis, we utilized SplitCap\footnote{SplitCap: https://www.netresec.com/?page=SplitCap} to segment the collected traces into individual TCP/UDP sessions. During this process, we discarded any unencrypted sessions or unrelated network conversations, such as DNS and ARP traffic, to maintain a focus on encrypted communications.

The remaining split sessions represent individual, encrypted connections that were established to request various resources from servers in order to load the 660 URLs. To provide meaningful labels for these encrypted TCP/UDP sessions, we employed a Server Name Indication (SNI)-based labeling strategy. This technique, inspired by practices from previous studies \cite{foundation_et_bert, dataset_sni_based_labelling_1, dataset_sni_based_labelling_2}, uses the SNI field within the TLS handshake to identify and label each session with the corresponding server name. This labeling approach ensures that each session is accurately associated with its originating domain, allowing for precise and reliable classification of network traffic based on the requested web resources.



\section{Evaluation - Data Preprocessing}
\label{app_sec:data_preprocessing}

To represent various design choices using ET-BERT and YaTC, we extracted raw data following the methods outlined in Table \ref{tab:8_pre_processing_techniques}. For both models, we adhered to the originally proposed number of bytes and packets to accurately capture different granularities. We also maintained the train, test, and validation splits as defined in the original studies to ensure consistency. 

For both ET-BERT and YaTC, data selection incorporates all layers—\textit{L2, L3, L4}, and \textit{L7}—unless stated otherwise.

\begin{table}[h]
\renewcommand{\arraystretch}{1.8}

    \caption{Data preprocessing for selected models}
    \label{tab:8_pre_processing_techniques}
    
    \centering
    \small

    \resizebox{\columnwidth}{!}{
        \rowcolors{2}{}{lightgray}        
        
        \begin{tabular}{llccc} 
            
            \toprule
            
                \multicolumn{1}{c}{\textbf{Model}} & 
                \multicolumn{1}{c}{\textbf{Packet}} & 
                \multicolumn{1}{c}{\textbf{T1}} & 
                \multicolumn{1}{c}{\textbf{T2}} & 
                \multicolumn{1}{c}{\textbf{Burst, T3}} \\ 
            
            \midrule
            
                ET-BERT  & 
                 128B & 
                 640B & 
                 640B of 5 pkt &
                 128B per pkt of 5pkt \\
                
                YaTC & 
                 1600B & 
                 1600B & 
                 1600B of 5 pkt &
                 320B per pkt of 5pkt \\
                
            \bottomrule

            \multicolumn{5}{l}{ 

                \begin{tabular}[l]{@{}l@{}}

                    \rowcolor{white}
                    
                    \normalsize \textbf{B}=Bytes;
                    \textbf{pkt}=Packets;
                    \textbf{T3}=First $m$ bytes per packet of $n$ Packets; \\[-3pt]
                    \normalsize \textbf{T1}=First $m$ bytes; 
                    \textbf{T2}=First $m$ bytes of $n$ packets; \hspace{0.2cm} \\
                \end{tabular}
                
            }
        
        \end{tabular}
    }
\end{table}

\end{document}